\documentclass[11pt]{article}
\usepackage{amsmath,epsf,psfig}

\begin{document}

\newcommand{\prtl}{\partial}
\newcommand{\la}{\left\langle}
\newcommand{\ra}{\right\rangle}
\newcommand{\dla}{\la \! \! \! \la}
\newcommand{\dra}{\ra \! \! \! \ra}
\newcommand{\we}{\widetilde}

\title{Barrier Softening near the onset of Non-Activated Transport in
Supercooled Liquids: Implications for Establishing Detailed Connection
between Thermodynamic and Kinetic Anomalies in Supercooled Liquids.}

\author{Vassiliy Lubchenko\footnote{Current Address: Department of
Chemistry, Massachusetts Institute of Technology, Cambridge, MA
02139.}
\hspace{.1mm} 
and Peter G. Wolynes \\ {\it \small Department of
Chemistry and Biochemistry, University of California at San Diego}, \\
{\it \small La Jolla, CA 92093-0371}}

\date{\today}
\maketitle

\begin{abstract}
According to the Random First Order Transition (RFOT) theory of
glasses, the barriers for activated dynamics in supercooled liquids
vanish as the temperature of a viscous liquid approaches the dynamical
transition temperature from below.  This occurs due to a decrease of
the surface tension between local meta-stable molecular arrangements
much like at a spinodal. The dynamical transition thus represents a
crossover from the low $T$ activated bevavior to a collisional
transport regime at high $T$. This barrier softening explains the
deviation of the relaxation times, as a function of temperature, from
the simple $\log \tau \propto 1/s_c$ dependence at the high viscosity
to a mode-mode coupling dominated result at lower viscosity.  By
calculating the barrier softening effects, the RFOT theory provides a
{\em unified} microscopic way to interpret structural relaxation data
for many distinct classes of structural glass formers over the
measured temperature range.  The theory also provides an unambiguous
procedure to determine the size of dynamically cooperative regions in
the presence of barrier renormalization effects using the experimental
temperature dependence of the relaxation times and the configurational
entropy data. We use the RFOT theory framework to discuss data for
tri-naphthyl benzene, salol, propanol and silica as representative
systems.

\end{abstract}

A unified picture of the dynamics of supercooled liquids has emerged
based on a theory of random first order transitions \cite{KW,KTW,XW}.
The mean field approaches to structural glasses exhibit two
transitions - one, an equilibrium transition at low temperature
($T_K$) corresponding to an entropy crisis of the type heralded by
Kauzmann decades ago \cite{Kauzmann} and a high temperature
transition, which is purely dynamic, at a temperature we call $T_A$
\cite{KW}.  The pure dynamic transition would be sharp for systems
with infinite range interactions but for real liquids with finite
range interactions it is actually a crossover that signals a change of
transport mechanism from gas-like collisional dynamics to activated
dynamics on an energy landscape.  The dynamical transition is
predicted generically via mode-mode coupling theories.  Random first
order transition theory analyzes motions in the energy landscape
dominated regime by going beyond mean field to treat entropic droplets
\cite{KTW}.  The creation of these droplets resembles nucleation at an
ordinary first order transition but instead of leading to decay of a
macroscopic metastable state droplet creation yields a mosaic of
long-lived local structures in the liquid.

For a long time the cooperative regions predicted by this theory
remained unobserved.  During the last decade, however, a variety of
experiments using nonlinear spectroscopies \cite{Spiess}, single
molecule techniques and scanning microscopies
\cite{Israeloff,CiceroneEdiger} have revealed the dynamic
heterogeneities that are intrinsic to this mosaic picture of the
deeply supercooled liquid.  Still more recently microscopic
calculations based on the RFOT have made quantitative predictions both
of the mean barriers \cite{XW} and the fluctuations of the barrier
from region to region that give rise to nonexponential decay of
measured time correlations \cite{XWbeta}.  These calculations give
results that agree reasonably with experiment, using no adjustable
parameters, thus providing a great deal of confidence in the
underlying theory.  The recent quantitative calculations expand around
the low temperature entropy crisis transition that is extrapolated to
occur at a temperature below those we can access without falling out
of equilibrium.  The purpose of this paper is to examine some of the
corrections to the theory that arise from proximity to the high
temperature dynamic transition at $T_A$.

Near to the entropy crisis transition, the RFOT theory predicts that
viscosity and structural relaxation times grow in a super-Arrhenius
manner that can be described by the Vogel-Tamann-Fulcher (VTF)
equation.  We have argued before that deviations from this law are to
be expected as one approaches the (mean field) dynamical transition
temperature from below \cite{XW}.  This is because the dynamical
transition temperature, as viewed from the amorphous solid side, is a
spinodal where there is a breakdown of the identity of the long-lived
mesoscale structures of the mosaic.  As we shall see the mosaic
actually loses its identity at a somewhat lower temperature than mean
field theory would predict.

In RFOT theory the approach to the spinodal from below leads to a
softening of the interfaces of the mosaic, so the droplets shrink
faster than expected as the temperature grows and the activation
barriers fall faster.  Equivalently said, the barriers initially grow
faster upon cooling below the dynamic transition temperature than the
VTF law determined by $T_K$ would predict.  The mode-softening effects
near a spinodal thus provide a natural explanation for the so-called
high and low temperature VTF fits that are sometimes used empirically
for the viscosity and relaxation times.

The observed deviations from strict VTF behavior have previously been
taken as evidence against the idea that configurational entropy drives
glassy dynamics in the activated regime.  Here we see the converse is
true - the observed deviations are evidence that entropic droplets are
the main mechanism of glassy motions and that RFOT theory describes
these excitations much better then the older Adam-Gibbs argument
\cite{AdamGibbs}, which makes no prediction of such mode softening
effects.  We will show that by merely adding one material dependent
parameter $T_A$, the temperature of the transition to landscape
dominated behavior, the predicted barriers fit the data quite well.
In addition the remaining parameters have the dependence on measured
configurational entropy predicted by microscopic calculations.  While
the comparison of theory with experiment makes the mode softening
quite evident, the calculations we present are not a complete theory
of the dynamical crossover region - the mode coupling slow down
predicted above $T_A$ doubtless occurs and will be reflected in
specific aspects of the form of time correlation function decays.  The
pure mode coupling results will be cutoff by droplet effects too
before a strict divergence occurs.  Thus the temperature at which
landscape effects start should not be taken as too precise.  In
addition, near $T_A$ droplets become so cheap in free energy terms
that they may directly contribute to the configurational entropy
\cite{EastwoodW} which consequently becomes ambiguous close to $T_A$.
The ambiguity of $s_c$ has been explored before both from a strictly
theoretical viewpoint \cite{KTW,EastwoodW} and in the more practical
context of how to use calorimetric data properly to infer transport
properties \cite{RichertAngell,MartinezAngell}.

The objectives of this paper are twofold. First of all, we
demonstrate, by an approximate but quantitative argument and by
comparing with experiment that there is indeed droplet-droplet
interface softening due to order parameter fluctuations near the onset
of activated transport, i.e. at temperatures when the liquid is only
marginally supercooled.  We thus explicitly calculate the structural
relaxation barrier height in the entire temperature range where such a
barrier has a meaning. Further, and more ambitiously, we outline a
framework of quantitative analysis in which the details of the kinetic
slow-down can be interpreted in {\em all} glassy materials. The
explicit calculations of the relaxation barriers given here allow one
to clearly distinguish between the low $T$ regime where the activated
methodology must be applied, and higher temperatures where
mode-coupling effects dominate, in spite of the presence of some
activated processes.  The temperature of the crossover between the two
behaviors is automatically deduced.  We argue lastly that the
mode-coupling phenomena are strongly affected by the entropic
droplets.  It is {\em not} a goal of this paper to provide for a
convenient fitting formula, but instead we describe a particular
fitting procedure, directly tied to quantitative theory, to clearly
discern the two distinct transport regimes in supercooled
melts. Nevertheless, we do end up discussing fitting issues in some
detail, which has made this article long if not at times tedious. One
unavoidably develops a sense of appreciation of the problems faced by
those who pioneered the already existing attempts at analyzing
relaxation data without using a particular theory as a basis, in spite
of much apparent variation in qualitative behavior among different
chemical systems.  In particular, we have come to the conclusion that
any global fit to the kinetic data that fails to take account of
thermodynamic data is dangerous because of the correlation of fitting
parameters.

The organization of this paper is as follows: in the next section we
review the entropic droplet picture, its microscopic implementation
using density functional theory and describe how the theory
quantitatively explains the phenomenology of both "fragile" and
"strong" liquids.  We then explore the mode softening expected near
$T_A$ presenting arguments using both scaling and variational ideas to
quantify precisely how the mode softening changes the mosaic structure
and the barriers.  Following this, the resulting expressions are used
to fit viscosity and relaxation time data for several specific systems
where parallel thermodynamic data are available to us, so the
softening affects can be tested. We will learn in which temperature
range activated rearrangements dominate molecular transport.  Remarks
of general relevance to fitting the temperature dependence of rates
will be made, using the microscopic picture uncovered by the RFOT
theory. Finally we comment briefly on how mode softening effects will
show up in experiments directly probing dynamic heterogeneity where
deviations from strict VTF bevavior have worried investigators.

\section{The Random First Order Transition theory at the onset
of activated transport in supercooled liquids.}

The Random First Order Transition (RFOT) theory of the structural
glass transition \cite{XW,KTW} has provided a microscopic
understanding of molecular motions in supercooled liquids, which we
briefly review in what follows. A supercooled melt is obtained upon
rapidly cooling the substance below its melting point, so that to
avoid crystallization - a transition into the lowest free energy state
available at these temperatures.  Tautologically, {\em each} portion
of the supercooled liquid is not in its true equilibrium state, but
remains in a disordered, {\em meta}stable state. In yet other words,
each molecule of the substance is caught in a trap. Hence, molecular
motions in such a supercooled system consist of groups of molecules
switching between different traps: the current metastable structure
containing the molecule must be destroyed while a new one created
around the molecule's locale. Alternatively said, molecular motions
imply that locally, one solution corresponding to a minimum of the
free energy functional (FEF) must be replaced by another. In order to
quantify these notions one should compute the spatial extent of these
local traps and the energetics of how one trap replaces another
locally, as well as the mechanism by which such replacement
occurs. This is what the RFOT theory does. Consider as an empirical
fact, that a compact region encompassing $N$ independently moving
molecular units has $e^{s_c N}$ ways to pack itself comfortably into a
metastable configuration at temperature $T$, where $s_c$ is the so
called configurational entropy, per independently moving molecular
unit (crudely speaking, this entropy is experimentally well estimated
by the excess of the entropy of the supercooled liquid over that of
the corresponding crystal, although differences of vibrational
properties may alter this relation quantitatively). In our earlier
language, $e^{s_c N}$ is the number of the free energy functional
solutions in a region containing $N$ moving units, or, more concisely,
``beads''. Determining the number of beads per molecule is an
important issue in testing the theory essentially involving the issue
of how to map the thermodynamics of a molecular fluid onto that of a
nearly equivalent fluid of spherical entities. This will be discussed
in great detail in the final section). A region of space that can be
identified by a single mean field solution is called a mosaic
cell. Trivially, the free energy of system of size $N$ that could {\em
freely} switch between any two minima, is $-T s_c N$. However, in
order to replace one solution by another, a nucleus of this new
solution must be created within the old solution and this resulting
entropic droplet would need to grow to replace the current molecular
arrangement locally. Creating an interface separating two low free
energy solutions must cost free energy, otherwise this new
configuration would be a solution on its own, contradicting the
original premise of having two distinct solutions in one region. The
system with a domain wall, representing the growing droplet, may well
resemble an external solution of the free energy functional but at a
higher energy or effective temperature than the ambient system. It
turns out \cite{KTW,XW} that this surface energy scales not with the
surface area of the droplet, but in a more complicated way, so that
the effective surface tension coefficient decreases for larger
droplets according to $\propto 1/\sqrt{r}$, where $r$ is the radius of
a (spherical) droplet.  This unusual scaling is result of there being
a multiplicity of distinct metastable molecular configurations in a
supercooled liquid. The interface between two given (and generally
{\it \`{a} priori} poorly matching) solutions can always be wetted by
other, better matching, solutions. Such wetting does not occur for
example at the solid-liquid interface in a regular crystallization
transition, because the crystalline molecular arrangement is equally
structurally different from {\em any} liquid packing, and therefore
results in a large mismatch energy upon contact between those two
phases.  In the case of a supecooled liquid, in contrast, there is a
near continuity of structures with a broadly varying degree of mutual
disagreement, thus always allowing to find solutions mediating the
costly narrow interface between two arbitrary minima of the free
energy functional. The wetting argument was first conceived by Villain
in the context of the random field Ising model (RFIM) \cite{Villain},
where the smooth landscape of the ferromagnetic spin-spin coupling is
frustrated by a random static field. Even though all compromises are
energetically inferior to the zero field translationally invariant
Ising free energy, there are lots of them!  (The analogy of the RFIM
with structural liquids fails in one respect in that the RFIM model
does not have the translationally invariant analog of the low energy
crystalline arrangement that many glass forming liquids could assume
if cooled down slowly enough).  In the RFIM, the surface tension
renormalization due to wetting was estimated by calculating an energy
gain of using a thicker interfacial region, made up by those wetting,
less mutually frustrated solutions, at the expense of increasing the
total area of the otherwise thin {\em bare} interfaces between two
arbitrary solutions. This argument can be repeated, almost verbatim,
for supercooled liquids \cite{KTW}, as long as the bare (unwetted)
interface between two arbitrary solutions can be regarded as thin
(i.e. comparable to the molecular length $a$). We will see shortly
that this amounts to neglecting, in a mean field fashion, order
parameter fluctuations at the temperature $T_A$ at which the
longlasting metastable arrangements in the liquid start to
persist. (This temperature also arises as a viscosity catastrophe in
the mode-mode coupling theory
\cite{Leutheusser,Gotze_MCT,Kirkpatrick}).  Obviously, neglecting
otder parameter magnitude fluctuations is a good approximation away
from $T_A$, and this assumption has lead to quantitative successes
that we now list: First, the RFOT is the only microscopic theory that
derives the barriers for structural $\alpha$-relaxation in supercooled
liquids \cite{XW}, using only the value of the heat capacity jump at
$T_g$ as the input. The theory also has explained non-exponentiality
of the $\alpha$-relaxation \cite{XWbeta} above $T_g$ and its
dependence on the liquid's fragility, with no additional
assumptions. Furthermore, it is the only existing theory that computes
the cooperative length scale of those structural rearrangements (5-6
molecular diameters at $T_g$), without using adjustable parameters and
that agrees with values found in many experiments
\cite{CiceroneEdiger,Ediger_hydro}.  Knowing the cooperative length
allows one to predict deviations from the Stokes-Einstein relationship
between liquid's viscosity and the particle's diffusion coefficient
\cite{XWhydro}, as confirmed by several measurements
\cite{CiceroneEdiger,Ediger_hydro}.  Finally, it turns out that this
length scale also dictates the spatial density of low energy
excitations in glasses, observed at cryogenic temperatures \cite{LW},
as well as the frequency of the Boson Peak and the thermal
conductivity plateau in amorphous solids at low temperatures
\cite{LW_BP}.  Before continuing our story, we feel we should comment
on the molecular nature of the droplet interface tension, which could
have so far appeared as a rather abstract concept to the reader. On
the contrary, this tension is not abstract but is quite real and
arises from stresses quantifying stretching (or contraction) of the
inter-molecular bonds that would ordinarily hold together a metastable
molecular configuration near a potential energy minimum but now mean
that locally the region is in the vicinity of a saddle
point. Therefore we always use word ``activated events'' to describe
structural rearrangements in the supecooled liquid in the usual,
chemical sense.  The presence of activated dynamics and anharmonic
deformations is one of the essential features of the RFOT theory,
which makes it distinct from theories including only the purely
collisional effects (those we discuss later in some detail in
connection with the mode-mode coupling theory).

The RFOT theory thus states that at temperatures sufficiently 
lower than $T_A$, the free energy due to the entropic droplet
formation $-\frac{4 \pi}{3} (r/a)^3 T s_c$ is opposed by the interface
formation surface energy $4 \pi \sigma(r) r^2$ to yield the
following droplet growth free energy profile:
\begin{equation}
F(r)= 4 \pi \sigma_K(r) r^2  -\frac{4 \pi}{3} (r/a)^3 T s_c,
\label{F_r}
\end{equation}
where $\sigma_K(r) \equiv \sigma_0 (a/r)^{1/2}$ is the radius
dependent surface tension coefficient, as expressed through the
surface tension coefficient at the molecular scale $a$ - bead size:
\begin{equation}
\sigma_0 = \frac{3}{4} (k_B T/a^2) \ln((a/d_L)^2/\pi e).
\label{sig_0}
\end{equation}
Here, $d_L \sim 0.1 a$ (for all substances) is the so called Lindemann
length \cite{Lindemann} - the size of thermal vibrations of a molecule at the
mechanical stability edge of a solid. This Lindemann length is a
central concept in the RFOT theory and in this paper, therefore we
will discuss it in some detail below (the origin of formula
(\ref{sig_0}) will also be discussed then).
\begin{figure}[tbh]
\centerline{\psfig{figure=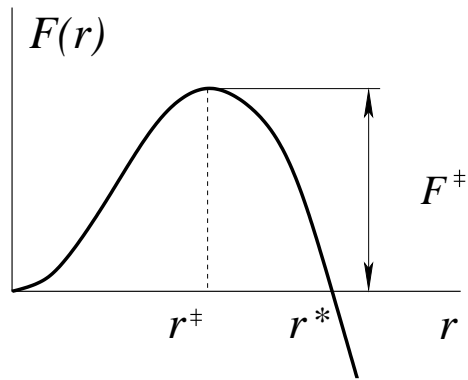,height=5cm}}
\caption{\small This schematic of the droplet growth free energy profile from
Eq.(\ref{F_r}) defines the critical radius $r^\ddagger$
($(dF/dr)_{r^\ddagger} = 0 $), the barrier height $F^\ddagger \equiv
F(r^\ddagger)$ and the radius $r^*$ of a spherical equilibrium mosaic
cell ($F(r^*)=0$, see text).  }
\label{F_r_fig}
\end{figure}
The RFOT theory indeed
makes many quantitative predictions, as briefly summarized above, but
perhaps more importantly it helps single out the universal aspects of
the glass transition physics in wildly different chemical systems that
superficially appear to have in common litle more than the rapidly
growing relaxation times as the temperature is lowered. We demonstrate
this by quoting some of the universal relationships found in
\cite{XW}, that help one to see the common molecular origin of the
glass transition in different systems. Indeed, combining Eqs.(\ref{F_r})
and (\ref{sig_0}) yields (see the schematic in Fig.\ref{F_r_fig}):
\begin{equation}
F^\ddagger/T = \frac{3 \pi [3/4 \ln(a^2/d_L^2 \pi e)]^2}{s_c} \simeq
\frac{32.}{s_c},
\label{F_sc}
\end{equation}
where $d_L/a \simeq 1/10$ is the (universal) Lindemann
ratio. According to Eq.(\ref{F_sc}), the extrpolated vanishing of the
configurational entropy (at some temperature $T_K > 0$), discovered
long ago by Kauzmann \cite{Kauzmann} would lead to divergence of the
relaxation barriers in a supercooled liquid at that very temperature!
The configurational entropy data are not readily available for many
substances, but the heat capacity jump $\Delta c_p$ is easier to
measure. If one approximates $s_c$ near $T_K$ in a linear fashion
according to $s_c \simeq \Delta c_p (T-T_K)/T_K$, one obtains that the
relaxation times should diverge according to the empirical
Vogel-Tamann-Fulcher (VTF) law: $\tau = \tau_0 e^{D T_K/(T-T_K)}$ and
automatically recovers a very simple relationship between the liquid's
fragility $D$ and heat capacity jump $\Delta c_p$ per independently
movable unit: $D = 32./\Delta c_p$ \cite{XW}. Thus, the RFOT provides
a constructive argument on the connection between the kinetic and
thermodynamical anomalies near the glass transition.  Note that in the
simple relationship in Eq.(\ref{F_sc}) the only temperature dependence
of the activation exponent comes from the $T$ dependence of the
configurational entropy. This is a consequence of the fact that the
droplet surface energy scales with the temperature itself, as shown in
\cite{XW} by a stability argument as applied to the liquid's FEF. The
value of the proportionality coefficient derived in \cite{XW} (leading
to the $32.$ factor above) must be correct within a factor of order
unity, however by good fortune it turns out to work very well in real
substances as is without any corrections for larger inteface width,
suggesting the unwetted interfaces of the mosaic cells are indeed
thin.  Now, one simple consequence of Eq.(\ref{F_sc}), is that the
magnitude of the configurational entropy per moving unit at the glass
transition is theoretically predicted to depend only on the magnitude
the relaxation time scale in the liquid:
\begin{equation}
s_c = \frac{32.}{\ln(\tau/\tau_0)}.
\label{sc_univ}
\end{equation}
On the other hand the glass transition temperature itself is defined
as the temperature at which the generic (equilibrium value!)
relaxation time in the system reaches a particular (large) value:
$\tau(T_g) = \tau_0 \exp(F^\ddagger/T_g)$. Therefore, the value of
$s_c$ (per bead) at $T_g$ depends only on the glass preparation time
scale, but not on the chemical nature of the substance!  For example,
if one patiently cools the liquid in a way that it is able to
equilibrate on one hour scale: $\tau/\tau_0 = 10^5 \, sec/10^{-12} \,
sec$, the resultant entropy is $s_c(T_g, 10^5 sec) = 32./\ln(10^{17})
= 0.82$.  Another common glass transition time scale - 100 seconds -
would yield $s_c(T_g, 100 sec)= 32./\ln(10^{17}) = 1.0$ Studying
available thermodynamical data indeed shows all substances exhibit
residual entropy of this magnitude at the glass transition (we will
discuss how to determine the number of moving units per molecule in
some detail later). We find it remarkable that a theory that uses no
assumptions other than the existence of the glass transition itself
(and uses no adjustable constants) should be able to provide the
magnitude of this residual entropy and explain why its value should be
so consistent among very different substances.  No other theory has so
far provided a microscopic basis for this consistency.  

This is perhaps a good point to make some comments about what the
essential assumptions in the RFOT droplet argument really are.  Other
than the surface tension $\sigma_0$ and the wetting effects that
decrease it, many of its ingredients are well established on empirical
grounds alone.  The $s_c$ measured by calorimetry can and should be
used in implementing the theory.  The Lindemann parameter (that goes
into determining $\sigma_0$) can be measured directly by neutrons.  It
is thus possible to remain quite agnostic about the ultimate vanishing
of $s_c$ at $T_K$ in applying RFOT theory to measurements. For the
purpose of the RFOT theory, $T_K$ is merely a convenient place to
estimate $\sigma_0$ because the free energy of a frozen minimum is
then decomposable into a localization part and an interaction (see
\cite{XW}), since the configurational entropy would vanish at $T_K$.
But if the comparison had to be made at the higher $T_g$, $\sigma_0$
would be smaller.  This diminution is essentially related to the
barrier softening dealt with in this paper.  The $r$-dependence of the
surface tension from the wetting requires a sufficient multiplicity of
states and nothing more. Again, a precise $T_K$ (referring to a strict
non-analyticity rather than a cross-over) is not needed to have this
effect. If a multiplicity of states exists, in any event, there will
be a strong reduction of the surface term from a naive surface energy
as the displaced region grows in size.  Deviations from the
$1/r^{1/2}$ behavior can be at most a factor of two in the size range
that can be explored to date in liquids without involving
supercosmological waiting times.  If indeed there is no crisp $T_K$
but rather a smooth turnover to an $s_c$ that strictly vanishes only
at absolute zero nothing about the predictions would be changed in the
temperature range that can be currently measured.

We point out to those scientists who, as a matter of faith, prefer a
smooth turnover in entropy rather than a sharp transition at an
entropy crisis that a complete theory in those terms requires in our
view a specification of what the $T=0$ state actually is supposed to
be and also how one should compute the excitation energies from it.
The only concerted attempt in that direction that we are aware of is
Nelson's theory \cite{Nelson} which, roughly speaking, suggests a
Frank-Kasper phase as the ground state and disclinations as the
defects.  Calculating the dynamics of disclinations at their
relatively high density above $T_g$ however remains still open in that
theoretical picture: the mere existence of such defects would in no
way be in direct contradiction to the present arguments.  Nussinov
\cite{Nussinov} has recently argued that the icosahedratic Hamiltonian has a
uniformly frustrated glass transition much like that predicted for
stripe phases \cite{SchmalianW} and therefore would be described by RFOT theory,
in so far as its dynamical behavior below $T_A$.

Another basic prediction of the RFOT theory has to do with the length
scales pertinent to the motional degrees of freedom in the activated
transport regime.  First of all, the critical radius for droplet
nucleation is, as easily shown from Eqs.(\ref{F_r})-(\ref{F_sc}),
\begin{equation}
r^\ddagger/a = (3 \sigma_0 a^2/2 T s_c)^{2/3} \simeq \frac{2.0}{s_c^{2/3}}, 
\label{r_bare}
\end{equation}
where $s_c$ is per moving unit. Hence, like the configurational
entropy, the critical radius depends only on the cooling rate at the
moment the liquid falls out of equilibrium. For example, at $\tau$'s
on an hour scale, $r^\ddagger/a \simeq 2.3$. This correspond to
reconfiguring about $48$ beads - a rather large (but mesoscopic)
number. While the volume of the critical transition state droplet
dictates the smallest number of molecules that needs to be summoned up
in order to convert a given region from one structural state to
another, the volume of the liquid that must ultimately reconfigure to
a state of a similar energy density is found from a slightly different
requirement. Indeed, the free energy profile for droplet rearrangment
from Eq.(\ref{F_r}), though very similar to a regular nucleation
theory, is quite distinct in the sense that the state on the other
side of the barrier is not a different phase of the substance, but is
really another liquid state!  To give an analogy, alternative liquid
(or, amorphous) packings are different in specific structural detail
yet are {\em generically} the same, just like fingerprints of
different individuals. Therefore at the droplet size $r^*$ determined
from $F(r^*)=0$, the liquid is guaranteed to be able to arrive to {\em
any} liquid state typical of this temperature. Thus to achieve an
irreversible arbitrary translation of an individual molecule will
require rearrangement of fluid within radius $r^*$. If a smaller
amount than $4 \pi/3 (r^\ddagger/a)^3$ is reorganized, the resultant
local energy density will be higher and the region is likely to revert
for its original configuration. On the other hand, if a region of
radius $r^\ddagger$ is rearranged, that rearrangement will continue to
expand until it extends a distance $r^*$, roughly. Eq.(\ref{F_r})
strictly would imply spherical droplet geometry but fluctuations from
such shape are likely because of the fractal wetting of the mosaic
cell interface \cite{Villain}. At any rate, we would like deduce a
{\em volume} density of the cooperatively rearranging molecular
structures; to this end we prefer to use a volumetric length scale
$\xi$, defined as $(\xi/a)^3 \equiv (4 \pi/3)
(r^*/a)^3$. Consequently,
\begin{equation}
\xi/a \simeq \frac{5.1}{s_c^{2/3}}.
\label{xi_bare}
\end{equation}
At the one hour scale glass transition $\xi$ is about $5.8$ moving
units. Cooperative lengths of this size have been observed in
experiments, as already mentioned. The number of those units in a
cooperative volume $(\xi/a)^3$ is about 200, but will be smaller at
higher temperatures, as is easily seen using Eq.(\ref{xi_bare}) when
we remember that $s_c$ grows with temperature. To summarize the idea
of cooperativity, in order for even a single molecule to perform a
translation, a large number of molecules must be moved as well,
otherwise the particle is likely to return to its original
location. The magnitude of each individual molecular translation is
about the Lindemann length $d_L$.  Such cooperative rearrangements are
intrinsic degrees of freedom in glasses that exist due to the
non-equilibrium nature of the glass transition. Their quantum
counterpart is the origin of the two level systems in glasses seen at
cryogenic temperatures \cite{LW}. The whole sample at any instant of
time may be imagined as a {\em mosaic} of individually cooperative
domains of roughly the size $\xi$ \cite{XW}.

Again, according ot the RFOT theory, the values of $s_c$ and $\xi/a$
at the glass transition mostly depend on the most probable relaxation
rate in the substance, but not much on chemical detail. While it has
been known for a while that the configurational entropy per molecular
unit is about the same for all substances at $T_g$, explicit
measurements of the cooperative length have been carrried out only
relatively recently, but they do recover a length of the order several
molecular spacings. Clearly, without a constructive argument for the
mechanism of the relaxation and the corresponding barrier height, the
existence of these universalities would thus be rather mysterious.
While here we see these universal patterns are quite natural, in
contrast the venerable Adam-Gibbs (AG) approach \cite{AdamGibbs} does
little to justify the correctly estimated $F^\ddagger \propto 1/s_c$
dependence, whose validity near $T_K$ we can now finally understand.
Indeed, the AG argument is not rigorously speaking a kinetics based
theory; instead in the AG paper, the estimates of the probability to
overturn a region was based on a detailed balance requirement,
something that specifies only {\em relative} rates.

Finally we note that the universal relationships from 
Eqs.(\ref{sc_univ})-(\ref{xi_bare}) will be somewhat modified when
the interface softening effects are taken into account. The
non-universal corrections will be especially pronounced
in the proximity of $T_A$, where the barriers for structural
rearrangements vanish. Conversely, substances with a smaller ratio
of $T_A$ to $T_K$ will exhibit larger deviations from universality
at the laboratory glass transition temperatures.

\subsection{Onset of non-activated dynamics at $T_A$: A temperature
driven Spinodal Transition between alternative metastable molecular
packings.}

\begin{figure}[tbh]
%\epsfysize=6cm
%\centerline{\epsfbox{k_k_theor_fig.eps}}
\centerline{\psfig{figure=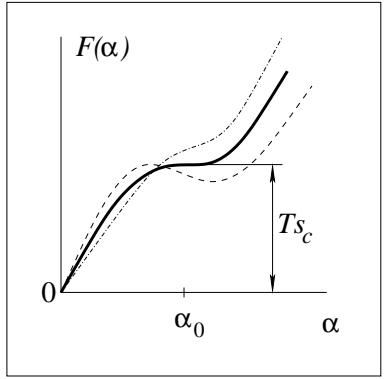,height=5cm}}
\caption{\small This is a schematic of the free energy density of an
aperiodic lattice as a function of the effective Einstein oscillator
force constant $\alpha$ ($\alpha$ is also an inverse square of the
localization length used as input in the density functional of the
liquid. Specifically, the shown curves characterize the system near
the dynamical transition at $T_A$, when a secondary, metastable
minimum in $f(\alpha)$ begins to appear as the temperature is
lowered.}
\label{F_alpha}
\end{figure}

From the liquid state viewpoint it makes sense to take a closer look
at what happens in the liquid when it is cooled below $T_A$, where the
barrier for structural rearrangements in liquids begins to appear. We
know of two different, but related ways to formally analyze phenomena
at $T_A$. One is the so called Mode-Mode coupling theory (MCT)
\cite{Leutheusser,Gotze_MCT,Kirkpatrick}, the other is the Density
Functional theory (DFT) \cite{SSW} or the nearly equivalent
self-consistent phonon argument \cite{SW}. A dynamical transition at a
temperature $T_A$, at which the liquid's viscosity diverges, was
discovered by using the MCT.  Although the result of an elaborate
self-consistent computation of the liquid's particle auto-correlation
function, this transition can be pictures microscopically as a
crowding phenomenon of a sort: molecules get in the way of each
other's motion, and at a certain temperature their thermal motion is
so slow that they cannot get out of each other's way and thus
effectively arrest their mutual displacement.  This dynamical arrest
appeared also, in what at first seems a little related manner, in the
DFT and the self-consistent phonon theory \cite{SSW,SW}. Those appear
to be {\em static} theories, which show that at low enough
temperatures, the molecules would give up their translational entropy
in favor of localization and forming aperiodic lattices! What would be
the form of the free energy functional that would let us detect such
singularity and what is the order parameter that would serve as the
argument of this functional?  We shall see an effectively useful
choice is quite simple. We reiterate that we are considering here a
localization transition below which a molecule settles down with its
surrouding in a metastable manner for times longer than the thermal
equilibration times, or several molecular vibrations, or, equivalently
at these temperatures, a few molecular mean passage times.  Such a
situation is not a theoretical speculation, but is the sensible
meaning of neutron scattering experiments that show a long plateau in
the time dependent structure factor \cite{Mezei}. At a first glance,
since the liquid (as empirically known from calorimetry) still has an
enormous number of alternative low density aperiodic structures at
temperatures where the plateau is seen, it seems one must have just as
many order parameters to describe completely all those alternative
outcomes after the transition takes place. This is directly related to
the necessity of introducing replica symmetry breaking (RSB) in mean
field spins glass models \cite{SGandB}). Yet, again, while in detail
each of these states is different, they do have common statistical
properties.  Describing the RSB is complicated enough in infinite
dimensions, but here we need to work with its consequences in three
dimensions.  There is a way to make quantitative progress, if one
realizes that the problem can be divided into two relatively separate
parts. Assessing the free energy change due to the topological, or
localization part of the transition can be separated from the
combinatorial component, which has to do with the multiplicity of the
alternative structures below $T_A$ corresponding to the same degree of
packing. The localization contribution was studied in \cite{SSW} by
calculating the free energy of a particular aperiodic structure
depending on the degree of localization as measured by the inverse of
the length scale $\alpha$ characterizing the spread of the molecule's
coordinate probability distribution: $\rho({\bf r}) \equiv \rho({\bf
r},\{{\bf r}_i\}) = \sum_i \left(\frac{\alpha}{\pi}\right)^{3/2}
e^{-\alpha ({\bf r}-{\bf r}_i)^2}$.  Clearly, $\alpha = 0$ describes a
completely delocalized liquid state, while $\alpha \rightarrow \infty$
corresponds to freezing into an infinitely rigid lattice. Therefore,
$\alpha$, if sufficiently large, can also be interpreted as the spring
constant of an Einstein oscillator restraining each molecule to its
location in the aperiodic lattice, hence the term self-consistent
phonon theory. In \cite{SSW}, the density profile $\rho(\{x_i\})$ was
used as a variational anzatz in the Ramakrishnan and Yussouff
\cite{RamakrishnanYussouff} density functional:
\begin{equation}
F[\rho({\bf r})] = \int d^{3}{\bf r}  \rho({\bf r}) [\ln \rho({\bf r})-1]
+ \frac{1}{2} \int \int d^{3}{\bf r} d^{3}{\bf r}' 
[\rho({\bf r}) - \rho_0] c({\bf r}, {\bf r}'; \rho_0) [\rho({\bf r}') - \rho_0]
+ F_{uni}.
\label{F_rho}
\end{equation} 
Here, $F_{uni}$ is the excess free energy of the uniform liquid, and
computing the interaction term requires knowing the direct correlation
function $c({\bf r}, {\bf r}'; \rho_0)$. $\rho_0$ is the average
density. Finally, we repeat, the set ${\bf r}_i$ correspond to the
coordinates of the vertices of a {\em particular} amorphous lattice.
Given the $\rho({\bf r})$ anzatz, $F(\alpha)$ can be computed
numerically for any lattice (the details of the calculation along with
the a discussion of the accuracy of the approximations used can be
found in \cite{SSW}). We note that other procedures have also lead to
the generation of aperiodic free energy minima on computers
\cite{DasguptaValls,PGWaper}. The main result is that now we have a free
energy as a function of a single averaged scalar order parameter
$\alpha$ that characterizes localization of the liquid's molecules
into amorphous packings. While at low densities a uniform liquid
situation $\alpha = 0$ is the only minimum of $F(\alpha)$, at some
high density/low temperature a secondary minimum appears at $\alpha$,
signalling a free energy non-analiticity. This is illustrated in
Fig.\ref{F_alpha}. The actual position $\alpha_0$ of this secondary
minimum depends on temperature; it is of the order $10$ at $T_A$, but
increases to become around $10^2$ closer to $T_g$.
Still, what can the {\em thermodynamical} significance of this near
spinodal minimum in so far it is much higher in free energy than the
delocalized state with $\alpha = 0$?  This significance may be hard to
appreciate until one recalls that even though each aperiodic
structure with $\alpha_0 \sim 10^2$ is indeed only {\em meta}stable
energetically, there are exponentially many alternative aperiodic
structures as indeed reflected in the large value of configurational
entropy.  Furthermore, the value of the free energy at the metastable
minimum is exactly equal to $T s_c$ at the corresponding temperature!
This has been rigorously shown for the Potts Glass, which also
exhibits a dynamical arrest at $T_A$ \cite{KW} and has similar
symmetry properties to the structural glasses. Recent explicit replica
approaches of structural glasses also make this point clear
\cite{MezardParisi}.

According to the definition of $\rho({\bf r})$, the inverse square
root of $\alpha$ corresponds to the localization length scale of a
molecule at $T_A$.  Alternatively, this length scale must be equal to
the amplitude of thermal vibrations of atoms around their metastable
equilibrium positions at the edge of mechanical stability of a
sample. The concept of this length scale was introduced long ago by
Lindemann \cite{Lindemann}. Empirically the Lindemann length $d_L$ is
known to be about one tenth of the molecular size $a$ from the
magnitude of the plateau measured in neutron scattering \cite{Mezei}. In
addition to the DFT and self-consistent phonon studies, later
calculations using more sophisticated, replica machinery
\cite{MezardParisi} confirmed the appropriateness of $\alpha$ as the
order parameter for the transition at $T_A$, as well as its numerical
value at the transition. The density functional theory explains the
appearance of the Lindemann ratio $d_L/a$ in the formula (\ref{sig_0})
for the molecular surface tension coeffcient between different
aperiodic structures. Though it was the result of a detailed
calculation using the free energy functional from Eq.(\ref{F_rho}),
formula (\ref{sig_0}) can be understood in simple terms by recalling
the expression for the free energy of a monatomic gas per particle:
$F_{ideal}/N = \frac{3}{2} \ln \left[ \left(\frac{eV}{N}\right)^{2/3}
\frac{mT}{2\pi \hbar^2}\right]$. The surface energy due to $\sigma_0$
is one half the free energy cost of the entropy loss of a particle
forced to be within a (localization) length $d_L$ instead of the
``legally'' allowed length $a$, corresponding to the liquid's volume
per molecule $V/N = a^3$ and relevant to interchanging particles'
identities. Lastly, we can already observe that due to the
fluctuations of the meanfield order parameter $\alpha$ at the spinodal
temperature, a sharp molecular interface can not be realized.
Moreover, due to the smallness of the barriers at the spinodal, which
we estimate in the next subsection, the surface tension between
alternative molecular packings will nearly vanish thus resulting in
very small droplets and extreme ease of their formation. This is what
we call {\em interface softening}.

The bare surface tension from the density functional argument for a
sharp interface between different aperiodic structures $\sigma_0 =
(3/4) \log(100/\pi e) (T/a^2) \simeq 1.85 (T/a^2)$ is actually quite
close to the simulated surface tension of a hard sphere crystal
approaching a hard wall \cite{HeniLowen}: $\sigma_{wall} \simeq 2.0
T_m/a^2$. The surface tension between a periodic crystal and its melt
is lower \cite{Laird}; $\sigma_{XM} = 0.61 T_m/a^2$ owing to the
broadness of the crystal-melt interface. We notice however that the
empirical validity of Turnbull's rule \cite{Turnbull} suggests that
the crystal/melt value is indeed universal. This is entirely analogous
to our assumption that $\sigma_0$ is universal and should vary little
from substance to substance.

Returning to the very beginning of this subsection, one may note that
it is not at first site obvious that the dynamical viscosity
catastrophe predicted by MCT and the DFT's localization transition at
$T_A$ should describe the same real life phenomena. Nevertheless, this
is indeed the case, as shown in \cite{KW_A}. That work exploited the
fact that the density functional allows one to derive both the
equilibrium value of the static order parameter $\alpha$, and the
molecular auto-correlation function, whose long time asymptotics
depends on $\alpha$ as well.  Moreover, in the infinite dimensional
limit the self-consistent equations determining the localizaton length
are identical in both descriptions. We mention this yet for another
reason.  Even though we will use the static formalism to estimate the
droplet surface tension close to $T_A$ and thus deduce the value of
the corresponding nucleation barriers, the dynamical perspective
supplied by the MCT will be necessary to achieve a full quantitative
description of the viscous phenomena at the very onset of activated
transport, where the crowding effects are primarily responsible for
the kinetic slowing down.  Conversely, we now understand that the
kinetics of these droplet rearrangements is influenced by the
smallness of the spinodal barriers in the vicinity of $T_A$, therefore
the analysis of the interface softening should be done
self-consistently. There is another, very important point we must
mention here. So far we have lead our discussion as in this paper if
the transition at $T_A$ occured as a sharp point in real
liquids. Actually, this transition is avoided due to appearance of
activated transport, or the entropic droplet excitations, as noted
long ago in \cite{KTW}. According to the MCT itself, at some
temperature ($T_A$) the liquid's viscosity would become {\em strictly}
infinite. Yet clearly molecular molecular motions that involve larger
fluctuations than the Gaussian fluctuations presented in the MCT are
possible but rather unlikely. These are the ``activated
events''. Hence, below MCT's $T_A$, the activated transport must begin
to dominate. In the context of liquid helium, Feynman once argued
\cite{Feynman} that the very definition of something being a liquid
means there can be no barriers.  In fact there is no contradiction
with his suggestion, because the barriers are between states that
correspond to quenched aperiodic solids structurally nearly identical
to the energy minima. A similar sentiment underlies Kuhn's theorem of
viscous polymer dynamics \cite{deGennes}.

\subsection{Estimate of the  barriers close to $T_A$.}

In this subsection we estimate the effects that interface softening
near $T_A$ will have on the value of the droplet surface tension,
assuming we know $T_A$ as an input parameter of the calculation.  The
approximate functional form we obtain will be then used to fit the
experimental $T$ dependences of the log-relaxation time for several
systems.

Estimating the surface tension betwen distinct phases can be done in
several ways. If a free energy functional describing all of the phases
of the system is known, there is a straightforward mean-field recipe
allowing one to complete the task (see for example \cite{Widom}), for
a given droplet geometry, or order parameter values at the boundaries
etc. This is done by integrating the free energy density, given by the
functional, throughout the interface region. Qualitatively, this
amouns to multiplying the functional's barrier height per unit volume
(it has the units of energy density!), computed using an appropriate
reference, by the interface width. At a second order transition, for
example, the latter is approximately equal to the (diverging)
correlation length.

Such a calculation close to $T_A$ can be done, in principle, by using
a computed functional $f(\alpha)$, allowing $\alpha$ to be a function
of $r$. One sticking point is that realistic, dense fiducial
structures characteristic of minimum free energy structure found near
$T_K$, cannot be easily generated with current simulation algorithms
in a reasonable time. One has to extrapolate to find their statistical
properties. The earlier RFOT theory calculations circumvent this
problem (in fact the closer to $T_K$, the better it should work) by
using the (measured) Lindemann ratio and assuming a sharp
interface. Near $T_A$, a more detailed free functional would
apparently have to be used.  Nevertheless, the surface tension near
$T_A$ can be estimated via a scaling argument.  Since both limits, at
$T_K$ and $T_A$, are under some control, an interpolated formula for
intermediate temperatures can be developed.

\begin{figure}[tbh]
%\epsfysize=6cm
%\centerline{\epsfbox{k_k_theor_fig.eps}}
\centerline{\psfig{figure=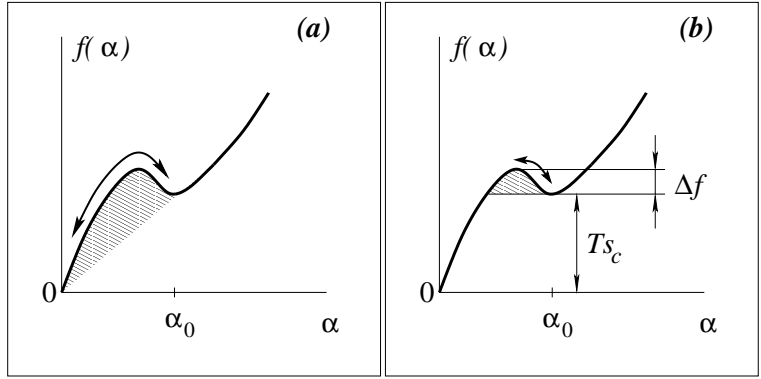,height=5cm}}
\caption{\small (a) The shaded area below the free energy curve shows
the free energy density contributing to the surface tension between an
aperiodic liquid packing and the uniform density, high temperature
liquid state. The double-ended arrow shows the end-points of a process
that would take the system from one state to the other.  In contrast,
transitions between two aperiodic structures, as shown in (b), do not
involve visiting the uniform state. As a result, a smaller amount of
energy (per unit volume) is contained in the droplet-droplet
interface.  }
\label{surf_area}
\end{figure}
Let us address the surface energy term in the high temperature regime
close to the spinodal at $T_A$.  Let us first compute the surface
tension at a regular spinodal transition occuring near a critical
point.  We must note, the issue of the interface energy between
alternative amorphous packings, as described by $f(\alpha)$, is
however different from a regular spinodal involving only two distinct
phases, in the following way. To determine the surface tension between
any of the amorphous structure and the delocalized uniform liquid
state, one must compute the ``contents'' of the underbarrier region,
as shown by the shaded area in Fig.\ref{surf_area}(a). The arrow above
the shaded area denotes a ``transition between'' a localized and the
uniform state. On the other hand, in order to calculate the
solution-solution interface energy, one must remember that $f(\alpha)$
is not the full energy density, but only a projection of the full
functional onto a one-dimensional surface, parametrized by $\alpha$. A
transition between two structurally unrelated states and back is shown
by a double-ended arrow now in Fig.\ref{surf_area}(b)). Clearly the
barrier for such a transition can not exceed $\Delta f$ (from
Fig.\ref{surf_area}(b)), because one can always find a sequence of
states of the liquid connecting two arbitrary metastable minima. The
highest energy state is the top of the barrier because this point is
accessible from any of the amorphous structures (and the uniform
state, of course). The droplet-droplet barrier energy is shown the
shaded region in Fig.\ref{surf_area}(b).$\Delta f$ gives therefore the
upper limit to the excess free energy density between two solutions
required to connect two different minima characteristic of the
temperature $T$. It, along with the width of strained region, is
needed to estimate the bare surface tension overall.

Using a generic analytical expression for a functional $f(\alpha)$
exhibiting a spinodal transition at $T_A$, we can estimate $\Delta f$
close to $T_A$. A number of simple functional forms for $f(\alpha)$
can be used that involve only a few constants.  We propose a form that
has as many constants as we can fix using the known theoretical
constraints.  Consider the following free energy density:
\begin{equation}
f(\alpha) = T_A s_c^A \left[1 + (1+c_1 t)
\left(\frac{\alpha}{\alpha_0} - 1\right)^3 
- c_1 t \left(\frac{\alpha}{\alpha_0} - 1\right) \right],
\label{F_a1}
\end{equation} 
where $s_c^A$ is the configurational entropy at $T_A$, $t \equiv
\frac{T_A-T}{T_A} > 0$ is the reduced temperature, and $c_1$ is some
yet unknown (positive) constant.  The functional in Eq.(\ref{F_a1})
already satisfies some constraints, known from \cite{SSW}. First, it
yields a spinodal transition at the value $\alpha = \alpha_0$ of the
order parameter, when $T=T_A$. Secondly, $f(0)=0$ - the liquid only
has the uniform component. Now call $\alpha_0(T)$ the position of the
metastable minimum of $f(\alpha)$ from Eq.(\ref{F_a1}) at any $T<T_A$
($\alpha_0(T_A) = \alpha_0$, of course).  Note, while the functional
in Eq.(\ref{F_a1}) does satisfy $F[\alpha_0(T_A)] = T_A s_c(T_A)$
automatically, it is too simple to reproduce the constraint
$F[\alpha_0(T)] \le T s_c(T)$ at an arbitrary temperature (for
example, the height of the inflection point of this $f(\alpha)$ is
temperature independent). Still, it has the expected physical feature
of exhibiting the spinodal type scaling at $T_A$ (such as the one
occuring in the Potts Glass at $T_A$). In view of what was just said,
we only need a qualitative estimate for constant $c_1$.  This is
easily obtained by requiring, for example, that $F[\alpha_0(T_K)] =
T_K s_c(T_K) = 0$. A simple calculation shows that $c_1 = 3/t_K$,
where $t_K \equiv (T_A-T_K)/T_A$, and therefore
\begin{equation}
\frac{\Delta f}{T_A s_c^A}  = 4 \left(\frac{t}{t_K}\right)^{3/2}
\frac{1}{\sqrt{1+3 (t/t_K)}}.
\label{DF1}
\end{equation}
The argument leading to Eq.(\ref{DF1}) is crude, but yields physically
reasonable results. It gives the expected scaling of the barriers with
temperature, and the absolute energy density scale is determined by
$T_A s_c(T_A)$ - the only available parameter with these
dimensions. The numerical coefficient is not very well determined but
it does turn out to work well enough to make clear a main point of
this paper, namely the existence of interface softening at $T_A$.

An elementary calculation that yields that the corresponding free
energy curvature at the minimum at $\alpha_0(T)$ gives a scaling
$t^{1/2}$, which leads to the correlation length scaling at the
transition $\xi_A \propto t^{-1/4}$.  These mean field scaling results
for an ordinary spinodal transition were obtained, of course, long ago
by Binder \cite{Binder} for a magnetic field driven spinodal
transition in an Ising spin system. Kirkpatrick and Wolynes identified
this length scale as of interest near $T_A $ a long time ago
\cite{KW}. This length scale $\xi_A$ is also the length scale
associated with replica theoretic and mode-coupling theories of
dynamical structural inhomogeneities in supercooled liquids
\cite{Glotzer}. Now, if it were not for the entropic droplet excitations
below $T_A$, our estimates for the surface energy at $T_A$ would
simply involve the excess free energy of a planar surface $\sigma_A
\sim \Delta f \xi_A$.  But at $T_A$ the barriers for structural
relaxations vanish (remember $\sigma(T_A) \propto \Delta f$). Since
infinitely small barriers imply infinitely small droplets, this means
that the correlations on the diverging length scale $\xi_A$ simply
cannot be established, because it costs no energy (and time) for the
material to reconfigure at these temperatures. In other words, the
interface of such a small droplet cannot be taken as a plane.
However, a droplet cannot be smaller than the molecular size $a$. This
length therefore is the appropriate length scale for the inhomogeneous
saddle point solution close enough to $T_A$. A better estimate for the
surface tension to compute the entropic droplet cost is, as a result,
$\sigma_A \sim \Delta f a$. On the other hand, there should be little
to no wetting in this regime. We repeat that this is because the order
parameter fluctuations exclude the possibility of having a narrow
droplet-droplet interface.  Alternatively said, wetting cannot take
place on length scales smaller than the $\xi_A$, that would be
appropriate in the absence of the entropic droplets.  Hence, the
surface energy term close close to $T_A$ should have a regular,
quadratic, scaling with the droplet's size (which in any event will
prove to be quite small).

We thus establish that the following droplet surface energy terms are
valid at the opposite ends of the temperature interval $T_K < T <
T_A$. At $T_K$, $\Sigma_K = 4 \pi \sigma_K(r) r^2 = 4 \pi \sigma_0 a^2
(r/a)^{3/2}$.  At $T_A$, on the other hand, $\Sigma_A = 4 \pi \sigma_A
r^2 = 4 \pi \Delta f a^3 (r/a)^{2}$, where $\Delta f$ given by
Eq.(\ref{DF1}) should be an adequate approximation. Note, at low
enough temperatures, the barrier estimate at $\Sigma_A$ will exceed
the $\Sigma_K$ term (wetted or not).  At high temperatures (closer to
$T_A$), the cheapest way for the system to reconfigure is to jump over
the low spinodal barriers (even though this process is somewhat
modified by the entropic dropets destroying long range correlations at
$T_A$). In this regime, the viscosity will in fact be determined not
by the activated kinetics since the individual molecular
configurations are short lived, but rather by the crowding effects of
the mode coupling theory.  However, as $T_K$ is approached, the mode
coupling relaxation times would diverge, and the spinodal barrier
becomes larger than that of the entropic droplet rearrangements
(wetted or not, actually). The entropic droplet rearrangements become
the main contributors to the molecular motions in the liquid.  A
simple interpolation formula, that allows the system to choose the
smallest available barrier, is provided by the following expression
for the droplet energy growth, for instance:
\begin{equation}
F(r)= \frac{\Sigma_K \Sigma_A}{\Sigma_K+\Sigma_A} - \frac{4 \pi}{3} r^3 T s_c.
\label{F_r_interp}
\end{equation}
Obviously, the ratio in the r.h.s. of Eq.(\ref{F_r_interp}) tends to
$\Sigma_K$, as $T \rightarrow T_K$, and vice versa, to $\Sigma_A$
around $T_A$. One should regard our argument as a variational
calculation for what would otherwise be a complex replica instanton
calculation. There is more than one way to make an interpolation of
this kind. We feel this ambiguity, however, is unimportant for the
following reason: We will be use the functional form of $F(r)$ in
Eq.(\ref{F_r_interp}) to fit the experimental $\log \tau$ vs. $T$ data
in the following section, using $T_A$ as the fitting parameter. Even
if this form of the surface term is not exactly correct, varying $T_A$
should be able to compensate for any inaccuracy to some
extent. Similar reasoning applies to any degree of uncertainty about
the exact functional form and value of the barrier at $T_A$ calculated
earlier in this section.  This can also be absorbed into the best
fitting value of $T_A$, provided the functional form in
Eq.(\ref{F_r_interp}) is flexible enough. We will see {\it \`{a}
posteriori} that it is flexible enough indeed, at the temperatures
where the liquids viscosity is due to the activated character of the
molecular transport, not the MCT effects, that is at the values of
viscosity around 10 poise and higher.  What is the meaning of the
value of $T_A$ that is obtained?  $T_A$ provides an estimate of the
temperature at which a soft crossover to the onset of activated
transport takes place.  Since this is not a sharp transition, it is
perhaps inappropriate to worry about the inaccuracy of $T_A$. We may
at least qualitatively compare ratios of $T_A$ to $T_K$ with detailed
microscopic calculations. Our main goal here is to show that the
phenomenon of interface softening is consistent with available
experimental data. And it is indeed, as we discuss in the following
section. Lastly, one should anticipate from Eq.(\ref{F_r_interp}) that
substances with a smaller value of the $T_A/T_K$ ratio will exhibit
more pronounced deviations from the simple $F^\ddagger \propto 1/s_c$
predicted neglecting the softening effects. It should be already
obvious from the discussion leading to Eq.(\ref{F_r_interp}) not only
that the surface tension will disappear at $T_A$, but also the RFOT
scaling of the barrier with inverse configurational entropy will be
replaced by a conventional $F^\ddagger \propto 1/s_c^2$ valid in the
absence of wetting near $T_A$. Such a result was obtained by
Kirkpatrick and Wolynes \cite{KW} and is described as the result of
replica instanton calculations for structural glasses by Parisi
\cite{Parisi} and by Takada and Wolynes for random heteropolymers
\cite{TakadaW}.

\section{Fitting and Interpreting Experimental Relaxation Times.}

In this final section of the article we discuss the implications that
mosaic interface softening has for the temperature dependences of the
apparent relaxation times in supecooled liquids.  The existing RFOT
theory calculations provide a complete microscopic picture of
activated transport in these systems, and a quantitative description
of the kinetics sufficiently close to $T_K$. Furthermore we will show
here that the RFOT theory can provide a constructive basis to analyze
the $\log \tau$ data for {\em all} compounds on the {\em same} footing
through nearly the entire range of liquid state conditions. Heretofore
such a universal analysis has usually been considered difficult for
the following reasons: On the one hand there is the obvious common
feature of a dramatic kinetic slowing down in all glassy
systems. Furthermore, it seems all known systems exhibit to some
degree two types of kinetics commonly known as structural, slow
$\alpha$-relaxation, that we focus on in this work, and faster
$\beta$-processes.  The slow $\alpha$-relaxation is normally
straightforward to separate from the $\beta$ processes and it exhibits
many similar features among different systems, most prominently a
rapid growth, that is often fitted with the VTF form. Since the work
of Adam-Gibbs \cite{AdamGibbs}, where $F^\ddagger \propto 1/s_c$ was
suggested, many purely empirical attempts have been made to establish
a connection between the Kauzmann's entropy crisis at $T_K$ and the
relaxation time divergence. But these attempts are stymied because,
unfortunately, the $\tau(T)$ dependences, while covering many orders
of magnitude, show, one may even say, qualitative differences in how
the grow with decreasing temperature when looked at very carefully.
For example, stronger substances tend to appear almost Arrhenius-like,
tempting an interpretation using a simple activated event as the
transport mechanism; fragile substances (like salol or TNB) exhibit a
pronounced turnover in $\log \tau(T)$ that is not entirely
satisfactorily fit by a single VTF law.  Still, perhaps with an
exception of polymeric melts \cite{RSN}, where detailed analyses pose
some problems (to be discussed below), an unambiguous connection
between the kinetics and thermodynamics of many glasses has been
established in several recent works
\cite{HuangMcKenna,MartinezAngell,Mohanty}. It seems beyond reasonable
doubt that the decrease in the configurational, or excess liquid
entropy with temperature is, at the least, qualitatively correlated
with the observed concurrent rapid growth of the relaxation times, as
supported by analyzing many substances. As we have already mentioned,
this correlation is predicted by the RFOT theory, which not only gives
a constructive argument in support of the relation $F^\ddagger \propto
1/s_c$ conjectured by Adam and Gibbs early on but also computed the
proportionality coefficient on a microscopic structural
basis. Experimentally, this latter relationship was convincingly
demonstrated to work well close to $T_g$ for several substances for
which both thermodynamical and kinetic data are available
\cite{RichertAngell}. As mentioned earlier, the RFOT theory predicts
that close to $T_K$, when $s_c \simeq \Delta c_p (T-T_K)/T_K$,
$F^\ddagger \propto 1/(T-T_K)$ and the VTF should describe well the
temperature dependence of the relaxation time magnitude.

While the VTF works globally well for many substances, some materials
seem to require more complicated functional forms over the measured
temperature range.  In those systems, the low $T$ part of the data
could still be fitted by the VTF, however the exact choice of which
part of the experimental curve to use in the fit introduces a
considerable degree of ambiguity. We know of several, unrelated
attempts to deal with this ambiguity. One is to make a simultaneous
plot of $1/F^\ddagger$ versus $s_c(T)$ as elegantly done by Richert
and Angell in \cite{RichertAngell} in the context of checking the AG
relationship (we feel this is the most direct way to see a correlation
between thermodynamics and kinetics in supercooled liquids, and find
the evidence demonstrated in \cite{RichertAngell} compelling). Another
approach is to plot $[d\log \tau/d(1/T)]^{-1/2}$ vs. $1/T$
\cite{Stickel} (a VTF functional form would appear as a straight line,
when graphed in this manner). This way of presenting the data
suggests, purely empirically, that at least three distinct regimes of
relaxation are observed in substances such as salol or TNB (from the
RFOT point of view, those regimes could correspond to the collision
and activation dominated transport and an intermediate
behavior). There have also been suggested functional forms fitting
$\log \tau$ vs. $T$ quite well in the measured range but that do not
even imply the existence of any thermodynamic singularity below the
glass transition temperature, but simply a powerlaw growth
($(T-T^*)^{8/3}$) of frustration limited domain reconfiguration energy
away from a second order-like transition at some temperature $T^* >
T_g$ \cite{Kivelson}.  In general, many attempts to fit the viscosity
data $\eta (T)$, including those mentioned above, generally assume an
activation type expression $\eta(T) = \eta_0 e^{f(T)}$, where $\lim_{T
\rightarrow \infty} f(T) \rightarrow 0$, and take both the prefactor
$\eta_0$ and the numerical constants in the exponent $f(T)$ as
adjustable parameters. Taking all those parameters as adjustable
without theoretical prejudice is the main difficulty in obtaining an
unambiguous interpretation.

Consider a simple VTF-like $\eta(T) = \eta_0 e^{D T_K/(T-T_K)}$ as an
example. Assume also, for the sake of argument, that there were no
(visually) satisfactory fit of experimental $\eta(T)$ data for a
particular substance with a {\em single} VTF law.  Let us then attempt
to fit only some low $T$ portion of the data with the VTF (certainly,
the RFOT theory says this should work close enough to
$T_K$). Experience shows that small portions of experimental $\eta(T)$
data are smooth and featureless enough to be easily overfitted with a
three-parameter VTF form. Furthermore, while smaller portions of the
$\eta(T)$ data are increasingly easier to fit, the resultant values of
the fitting parameters attain increasingly more unphysical
values. Most notably, $\eta_0$, bearing the meaning of the high
temperature viscosity, will dramatically differ from its empirical
value, known to be of the order a centipoise for {\em all} substances
near their boiling temperature.  Consequently, in our view, one should
not trust the corresponding values of $T_K$ and $D$ obtained in such a
fit. It is our impression that most (but not all) workers that have
used the concepts of fragility and $T_K$ as fitting parameters are
aware of this point. Some other ways of analyzing the relaxation
times, such as using the concept of fragility index
\cite{mAngell,mNgai} partially avoid dealing with the prefactor
$\eta_0$ explicitly, but have the potential to hide an unphysical
$\eta_0$. On the other hand, if the value of $T_K$ and the
preexponential factor $\eta_0$ are independently known, the resultant
fits, although not visually perfect will yield more physical outcomes.
As far as the value of $\eta_0$ is concerned, the RFOT theory is
specific in that it requires that in the predicted $\eta(T)$ from
$\tau(T) = \tau_0 \exp(32./s_c)$, $\tau_0$ be equal to the mean
molecular flight time, or equivalently at these temperatures, a few
molecular vibrational times. As a result, the RFOT prescribes that the
{\em activated} contribution to the viscosity should equal $\eta(T) =
\eta_0 \exp(32./s_c)$, where $\eta_0$ is indeed the high
temperature/low density viscosity of the liquid, which could be
measured experimentally on short time scales or calculated by a
perturbative extension of the standard kinetic theory of a dilute gas
in the manner of Enskog.  Clearly, since the activation barriers at
$T_A$ vanish, the main source of molecular drag near $T_A$ will come
from the multiple collision dynamics described by the Mode Coupling
Theory and which is not addressed in this paper.  Still, activated
transport dominates at low temperatures, when the viscosity exceeds a
few poise for simple liquids. This number would need to be modified
for polymers where we know long relaxation times of the chain can come
in even for dilute solutions owing to the Rouse modes
\cite{deGennes}. The discussion above suggests that we employ the
following fitting procedure that is free of the overfitting problems
just mentioned and that only uses the RFOT predictions in the regime
of their presumed applicability. Suppose we have experimental
viscosity and configurational data for a substance between its glass
transition and the boiling temperatures. The activation contribution
to the viscosity is presented as $\eta(T) \equiv \eta_0
\exp(F^\ddagger/T)$, where the barrier $F^\ddagger$ is computed using
Eq.(\ref{F_r_interp}), using the experimentally observed $s_c$ and
some value of $T_A$, that in the end will be our fitting
parameter. The value of $\eta_0$ is not computed by the RFOT theory
but is known to be within a factor of two or so to to equal the high
temperature value of viscosity, although in fact it should be somewhat
density dependent owing to the collision rate dependence on
density. Isochoric measurements of transport properties are much
needed and should be easier to interpret! To avoid ambiguity, we will
take $\eta_0$ equal exactly the viscosity of the liquid at the boiling
point. The slight looseness of this procedure can only result in a
logarithmic error to the exponent itself, that we focus on in this
work. Recall that the $s_c$ in formula (\ref{F_sc}) is measured per
independently moving molecular unit (or, more concisely, per one
``bead''). The bead count will be presented for all four substances we
consider in this paper, however we can be sure of its exact value only
within say 30\% accuracy, with the degree of uncertainty being smaller
for simple liquids and larger for bigger molecules as we will discuss
below. Empirically, the temperature dependence of experimentally
inferred configurational entropy follows very well the following
functional form \cite{RichertAngell}: $s_c^{exp} = s_{\infty}
(1-T_K/T)$. (Here, the experimental input $s_c^{exp}$ is indicated per
mole, not bead!). Obviously, the parameter $s_{\infty}$ would be equal
to the heat capacity jump at the ideal glass transition temperature
$T_K$, while at the arbitrary temperature $T$ the jump is $\Delta
c_p(T) = s_\infty (T_K/T)$.  We will not fit with respect to $T_K$,
but take it from the experimentally determined $s_c(T)$ to avoid the
overfitting mentioned earlier.  However, since $s_\infty$ is known
directly only on a per mole basis, we find it convenient to treat the
overall factor for $s_c$ in Eq.(\ref{F_r_interp}) (remember this one
is per bead!) as a fitting parameter. We call this parameter
$s_{fit}$, it is therefore the fitted prefactor of the configurational
entropy per bead: $s_c(T) \equiv s_{fit} (1 - T_K/T)$. Now, if we knew
exactly the number of beads to be assigned per molecule (which we call
$b_K$), we could then directly compare quantity $s_\infty/b_K$ with
$s_{fit}$ in order to assess the quality of our approximations. The
ratio $(s_\infty/b_K)/s_{fit}$, reflecting the degree of deviation of
the theory from the experiment can be presented more conveniently as
$(s_\infty/s_{fit})/b_K$ - that is as the ratio of the bead count
$s_\infty/s_{fit}$ predicted by the RFOT theory (witihin current
approximations!) and our {\it \`{a} priori} bead count, to be
discussed below, in due time.  In an ideal situation, where the bead
count approach is exact, the value of $(s_\infty/s_{fit})/b$ would be
equal to one.  We should mention that in our minds the ``bead''
concept does not refer to some arcane renormalized object but merely
recognizes that almost all the substances treated here are made up of
aspherical molecules not hard spheres, so the calculation of the
interface tension refers to the localization of the quasispherical
parts of each such molecule. It would be much clearer, if RFOT could
be discussed in the context of molten salts or other mixtures of
spherical components, but all the relevant data for a test do not seem
to be available for these systems.  An independent theoretical
soundness check is that both the fitted and the experimental value of
$s_c$ per bead near $T_g$ should be on the order of the universal
$.82$ predicted by the vanilla version of the RFOT theory for all
substances at a glass transition on one hour time scale (although a
deviation is expected, and more so for substances with $T_A$ and $T_K$
closer to each other).  A deviation of the quantity
$s_\infty^b/s_{fit}$ from unity could, in principle, happen for four
(mostly independent) reasons: first, a possible inaccuracy of the RFOT
estimate of the droplet surface tension owing to a more complicated
interface structure; second, an incorrectly estimated bead count (the
number of beads per molecule of the substance); third, the approximate
nature of the argument leading to Eq.(\ref{F_r_interp}); and final
fourth, an uncertainty in experimental estimation of the
configurational entropy $s_c$ owing to different vibrational
properties of crystals and aperiodic structures (this is discussed in
some detail in \cite{RichertAngell}, see also \cite{MartinezAngell}).
We have already commented on the approximations of the softening
anzatz from Eq.(\ref{F_r_interp}), however we would like to say a few
words in connection with the first two possible sources of error. The
determination of the factor ``32.''  in Eq.(\ref{F_sc}) for simple
liquids uses no hidden assumptions and would be expected to be
accurate within no worse than a factor of two (see
\cite{XW}). Moreover this constant should be nearly the same for all
those substances owing to the near universal Lindemann parameter. At
the same time, polymeric molecules may exhibit effects of a finite
chain persistence length which is often hard to determine.  Clearly,
close enough to $T_K$ the persistence length will be smaller than any
relevant length scales in the problem, however just above $T_g$ the
critical droplet radius following from Eq.(\ref{F_r}) is still only
$2.3 \, a$ \cite{XW}, suggesting a possibility of microscopic
effects. Physically, persistence length phenomena will lead to a
larger effective bead size and thus are hard to separate from the bead
count issue. Additionally, polymeric substances often have beads of
very different size (see below), therefore using a single generic bead
size $a$ probably introduces another source of quantitative
discrepancy. More importantly, polymers may exhibit local ordering
transitions, such as crystallization or orientational ordering that
fail to grow to detectible size because of entanglement.  These
effects are clearly not included in our present analysis, not to
mention the effects of the existence of very long time relaxation
processes in polymer melts even in the absence of supecooling. A
separate theory for these effects is needed to achieve a quantitative
description of relaxation in liquids comprised of very long
molecules. To finish the discussion of the choice of fitting
parameters, we mention another reason why we use the entropy
coefficient as a fitting parameter rather than, say, the fragility $D$
from the VTF. Due to the empirical temperature dependence of the
configurational entropy - $s_c \propto (1-T_K/T)$ - a plain AG law
does not precisely follow the appropriate functional form even in the
absence of softening. Instead, the bare RFOT predicts a slightly
different $\log(\eta/\eta_0) = D_1 T/(T-T_K)$, as compared to the
VTF's $\log(\eta/\eta_0) = D T_K/(T-T_K)$. While asymptotically these
two forms are equivalent as $T \rightarrow T_K$, the exact numerical
value of the ``fragility'' factor $D_1$ will differ from the standard
fragility $D$ at $T_g$ by a factor of $T_K/T_g$, which can be
significantly different from unity ($\sim 1/2$ in the case of
a-Silica).

\begin{figure}[tbh]
\centerline{\psfig{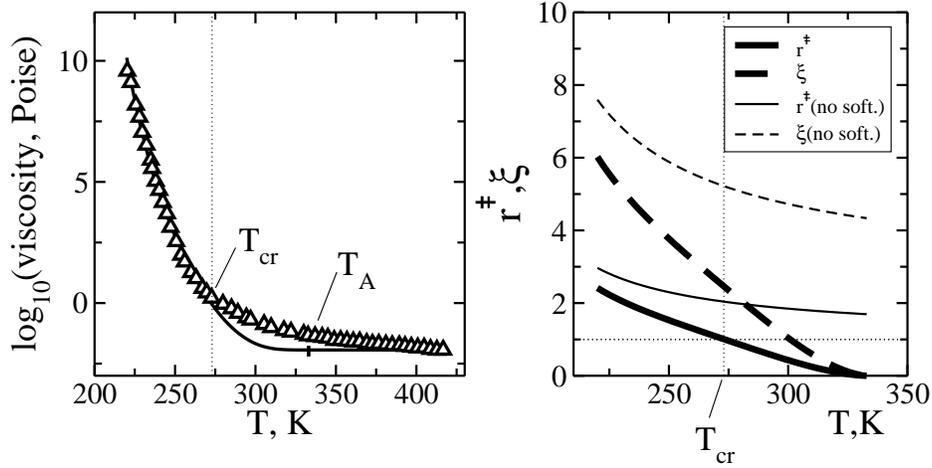}}
\caption{\small Experimental data (symbols) for salol's viscosity
(Ref.\cite{Stickel2}), superimposed on the results of our fitting
procedure (line) are shown. $T_A$ is shown by a tickmark.  The r.h.s.
pane depicts the temperature dependence of the length scales of
cooperative motions in the liquid. The thick solid and dashed lines
are $r^\ddagger$ and $\xi$ respectively (see text). The thinner
counterparts show what those lengths would be in the bare RFOT, but
calculated using the same value $s_{fit}$ as the renormalized lengths
(this comparison is only of limited value, because, remember that the
value of $s_{fit}$ that should be used to calculate $r^\ddagger$ and
$\xi$ in the absence of softening would be different from the one
computed with softening. Experience shows, the values of the length at
$T_g$, unlike $s_c(T_g)$, is rather insentive to whether softening is
present or not).}
\label{Salol}
\end{figure}

\begin{figure}[tbh]
\centerline{\psfig{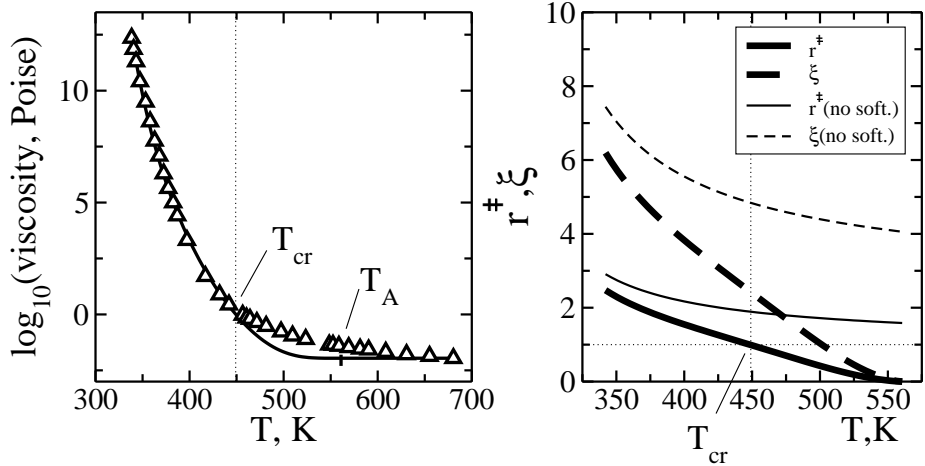}}
\caption{\small Same as Fig.\ref{Salol}, but for TNB. Experimental
data are taken from Ref.\cite{TNBexp}.}
\label{TNB}
\end{figure}

\begin{figure}[tbh]
\centerline{\psfig{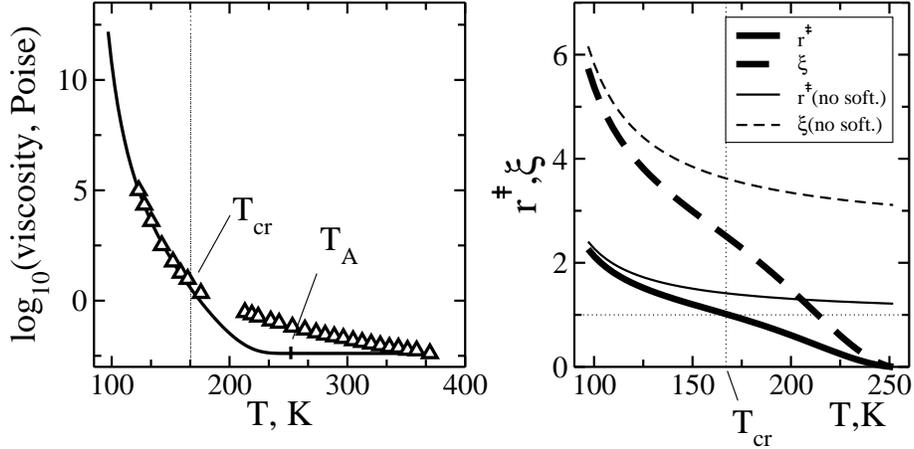}}
\caption{\small Same as Fig.\ref{Salol}, but for
propanol. Experimental data are taken from Ref.\cite{Stickel2}. Note,
viscosity points for this substance are available in a more narrow
range than, for example dielectric relaxation; in fact the data stop
at a temperature significantly above the conventional
$T_g$. Nevertheless, we prefer to use the viscosity measurements do
reduce ambiguity in the prefactor (see text).  The fitted
$\log(\tau(T))$ in the region of no experimental data should in fact
be considered as a prediction of the RFOT theory.}
\label{Propanol}
\end{figure}

\begin{figure}[tbh]
\centerline{\psfig{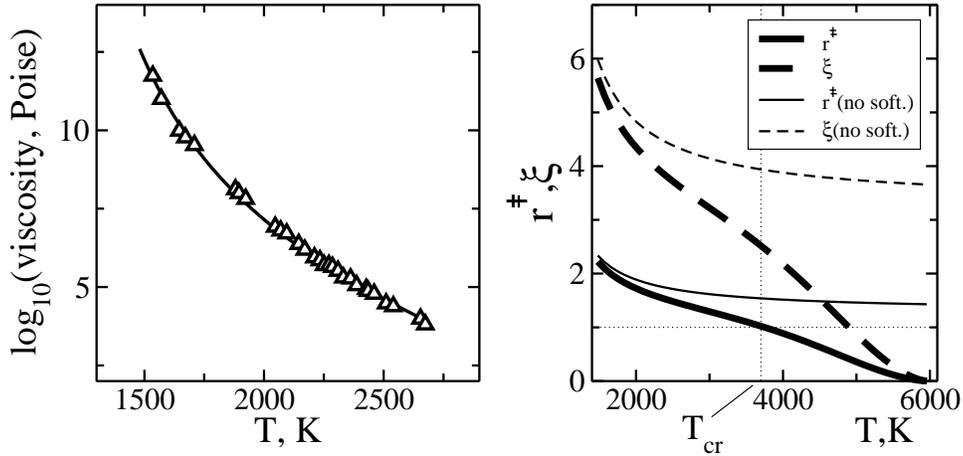}}
\caption{\small Same as Fig.\ref{Salol}, but for amorphous silica.
Experimental data are taken from Ref.\cite{Richet}. The value of $T_A$
obtained in our fit is too large to appear on this graph. It is in
fact higher than the boiling point of $Si O_2$.  This is a clear
indication that this substance is highly networked. Silica's $T_{cr} \simeq 3700 K$. }
\label{Silica}
\end{figure}

We have chosen four substances - silica, 1-propanol, tri-naphthyl
benzene (TNB) and salol - to demonstrate our theoretical
conclusions. This particular choice was suggested by the following
rationale. First, thermodynamic data are available for these
materials; second, these substances cover a wide range of fragility
values, with silica being the strongest substance, TNB and salol the
most fragile and propanol being in the intermediate range. From an
RFOT theoretical perspective TNB and salol are also attractive because
of their being notorious for deviating from the plain VTF dependence;
besides, the length scale of RFOT theory has been checked for TNB
\cite{Ediger_hydro}. The experimental points and the fitted curves are
shown in the l.h.s. panes of Figs.\ref{Salol}-\ref{Silica}.  The
corresponding parameter values are shown in Table \ref{parmtr_table}.
\begin{table}
\small
\centerline{\begin{tabular}{|c||c|c|c|c|c|c|c|c|c|c|} \hline
substance  & $T_g$,K & $T_K$,K & $s_{\infty}$,$k_B$ & $\eta_0$,cPs & $T_m$ 
& $s_{\infty}(T_m \! - \! T_K)/T_m$& $\Delta H/T_m$& $T_A$,K & $s_{fit}$,$k_B$ & $s_{\infty}/s_{fit}$
\\ \hline \hline
Salol      & 220  & 175  & 16.7 & 1.16 & 315 & 7.4 & 7.34  & 333 & 2.65 & 6.29
\\ \hline
TNB        & 342  & 271  & 25.14 & 1.1 & 472 & 10.7 & 10.8 & 561 & 2.68 & 9.37
\\ \hline
1-Propanol & 97.0 & 72.2 & 8.53 & .41  & 146 & 4.31 & 4.87 & 252 & 2.89 & 2.95
\\ \hline
a-Silica   & 1480 & 876  & 1.49 & 1.0 & 1700 & .72  & 0.66 & 5960 & 1.90 & 0.78
\\ \hline  
\end{tabular}}
\caption{\small First five columns give known quantities
(Ref.\cite{RichertAngell} for salol and propanol; Ref.\cite{Magill}
for TNB, Ref.\cite{Richet} for a-Silica). Columns six and seven
demonstrate consistency between the used form of the configurational
entropy (provided by other workers, as cited) and the known value of
the enthalpy of fusion. Indeed, by definition of configurational
entropy ($s_c = s_{liq} - s_{cryst}$): $\int_{T_K}^{T_m} (\Delta
c_p/T) dT = \Delta S_m \Rightarrow s_{\infty}(T_m - T_K)/T_m = \Delta
H/T_m$, if one uses the empirical form $s_c = s_{\infty} (1 -
T_K/T)$. Additionally, these numbers give an idea how many independent
moving units are in each molecule of the substance, if one adopts the
view that a fixed amount of entropy per such moving unit is lost at
freezing (see the extended discussion in the last section). Next two
columns show the fitted values of $T_A$ and $s_{fit}$
respectively. The ratio $s_{\infty}/s_{fit}$ is a quantity that,
within RFOT, should give the effective number of beads per molecule,
as expressed by the rate of entropy loss at $T_K$. }
\label{parmtr_table}
\end{table}

The r.h.s. panes of Figs. \ref{Salol}-\ref{Silica} show the
temperature dependence of the relevant length scales for each
respective substance. The thicker lines denote the critical radius
$r^\ddagger$ (solid) and the cooperativity length $\xi$ (dashed) in
units of molecular length $a$. The corresponding thin lines show what
those length scales would be in the bare RFOT theory without softening
$r^\ddagger/a$ from Eq.(\ref{r_bare}) and $\xi/a$ from
Eq.(\ref{xi_bare}).  Note, the configurational entropy, used to
compute the length scales with and without softening are taken from
the fitted value of $s_{fit}$, calculated {\em with} softening, to
allow for direct comparison, that is to illustrate: ``smaller
barriers'' $\Rightarrow$ ``smaller droplets''. Fitting the viscosity
data with a plain VTF would result in a slightly different bare value
of $r^\ddagger$ and $\xi$. In fact, while the precise $s_c$ at $T_g$
is sensitive to whether softening effects are included, the length
scales near $T_g$ (but not immediately near $T_A$) are not very
sensitive (we make comments in this later on).

We remind the reader that $T_K$ and $\eta_0$ were fixed by
independently measured thermodynamics and the measured viscosity near
boiling respectively; on the other hand $T_A$ and $s_{fit}$ were
varied to fit the low temperature part of the $\eta(T)$ curves
only. Since $T_K$ is not varied (from its thermodynamic value), the
outcome of the fitting procedure is rather insensitive to the exact
choice of the low $T$ fragment of the data used for fitting.  For
clarity, we used all experimental viscosity data above 10 poise in the
fit (and at $T > T_g$).  This is internally consistent within the
theory, as will be shown soon when analyzing the derived critical
droplet sizes (those should be at least one molecular length
$a$). Using fewer high $T$ points results in a somewhat higher value
of $T_A$, but the change in its value is not dramatic.  Lastly, in the
case of Silica, the high $T$ value of viscosity at the vaporization
point is not known (surprisingly, its boiling point itself is not
known well either), therefore we took (admittedly arbitrarily)
$\eta_0$(a-SiO$_2)= 1$ centipoise.

Let us inspect the fitted theory versus experiment graphs in
Figs.\ref{Salol}-\ref{Silica}. With the exception of silica, whose
data simply do not extend into high enough temperatures, all figures
demonstrate that there exists temperature ($T_A$) above which the
barrier for activated motions in the liquid disappears.  This is the
temperature at which the fitted log-viscosity begins to level off. We
remind the reader that only the activated part of the viscosity has
been computed in this work. Clearly, this activated mechanism becomes
subdominant at viscosities below 10 poise or so, where the
experimental and the curves fitted to the low temperature data begin
to diverge. With remarkable consistency among all substances
considered, this divergence temperature (call it $T_{cr}$) coincides
with the temperature at which the critical (transition state) droplet
radius equals the molecular size $a$ (see the r.h.s. panes of
Figs.\ref{Salol}-\ref{Silica}). We see therefore the RFOT theory
provides an internal criterion for when it should fail to
quantitatively account for the relaxation rate values - obviously, the
critical droplet size cannot be smaller than a molecule's size (within
a factor of order unity, of course; this factor, encouragingly, is
consistently around unity among the few materials analyzed here.  To
avoid ambiguity, we define $T_{cr}$ as the temperature at which the
value of the critical radius $r^\ddagger$ is strictly equal to the
molecular length scale $a$, although other conventions are possible).
While the equations described here do give results between $T_A$ and
$T_{cr}$, clearly a continuum treatment in this size regime is
suspicious since $r^\ddagger$ is less than a particle size. It makes
sense in this temperature range to invoke a thoroughgoing single
particle picture of the activated events.  Very promising work by
Schweizer and Saltzman \cite{Schweizer} along these lines appeared
just as this paper was finished. They use a quantity similar in spirit
to our $\Delta f$ to estimate the barrier for a single particle to
escape its cage, which should be adequate in this regime.

Our result that the crossover between activated and collisional
transport should take place at a specific time scale or viscosity is
consistent with recent experimental findings in \cite{Roland} on the
pressure dependence of relaxation times in several materials, where
there was observed a crossover between two types of the pressure
dependence of the dielectric response, similar to what is seen in the
$\log(\eta(T))$. As expected, at higher temperatures this crossover
occurs at higher pressures. Remarkably though, the crossover takes
place at the same value of the relaxation time!  This is a clear
indication that this transition has to do with excluded volume
effects, consistent with our conclusions that for $T > T_A$ the
collisional viscosity mechanism dominates.  The temperature $T_{cr}$
of the crossover between collisional and activated transport
corresponds well to the purely empirically defined break temperature
$T_B$, at which the Stickel's \cite{Stickel} $[d\log
\tau/d(1/T)]^{-1/2}$ vs. $1/T$ plot exhibits a visible kink (see also
\cite{Ngai_rev}). We have noted in the previous paragraph that there
is a degree of uncertainty about the value of the prefactor $\eta_0$
for Silica, whose $T_A$ is rather high which indeed turns out to be
above its boiling point. Obviously, quite an extrapolation is
involved. Nevertheless, we note that the crossover temperature
$T_{cr}$ at which our theory predicts a crossover between activated
and collision dominated transport mechanisms is about 3700 K for
silica. This is consistent with the simulations of Horbach and Kob
\cite{HorbachKob}, who recover a transition to dynamics characteristic
of fragile liquids at a temperature $T_c = 3330$.

As we have just seen, taking into account droplet interface softening
reveals that the critical transition state droplet size $r^\ddagger$
is smaller than expected otherwise from the bare version of the RFOT.
This is demonstrated in the r.h.s. graphs of
Figs.\ref{Salol}-\ref{Silica}.  The effect is more pronounced for
substances with a smaller value of $T_A/T_K$ ratio. This ratio, as we
have mentioned earlier, gives the relative value of the relevant
energy scales at the dynamical and ideal glass transition. The fitted
value of $T_A/T_K$ for the two most fragile substances is about $2.$;
the intermediate fragility substance has a fitted $T_A/T_K$ of $3.5$
and silica a value of $6.8$. According to the work of Hall and Wolynes
\cite{HW}, this ratio reflects the degree of network formation in the
substance. The Hall-Wolynes prediction for interacting spheres with a
pure $r^{-12}$ repulsion gives $T_A/T_K = 1.3$ while the ratio is
predicted to be much larger for silica, but of course the effect of
vaporization was not included in that work.  A relative decrease of
the softened critical radii is a consequence of the barrier vanishing
at the dynamical transition temperature $T_A$, as explained in the
previous section.

\begin{figure}[tbh]
\centerline{\psfig{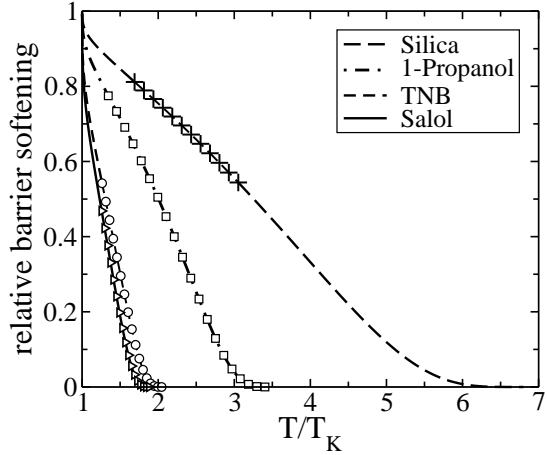}}
\caption{Shown here are the ratios of the softened relaxation 
barriers (from Eq.(\ref{F_r_interp})) to their bare values, calculated 
using Eq.(\ref{F_r}) and $s_c = s_{fit} (1-T_K/T)$.}
\label{rnrm}
\end{figure}

Let us now see quantitatively how the barrier height is lowered when
mode softening is included in the theory.  The ratio of the
renormalized barrier to its $T_K$-asymptotic form $F^\ddagger/T =
32./s_c$, calculated in \cite{XW}, is shown in Fig.\ref{rnrm}.
The graph in Fig.\ref{rnrm} is designed to show how the prediction of
the $T_K$-asymptotic RFOT theory is modified when the barrier
vanishing at the spinodal $T_A$ is included in the treatment.  This
more complete theory, albeit approximate, explicitly shows that the
barriers for droplet rearrangements must disappear at some
temperature, that we have called $T_A$. Of course, since the critical
droplet size is smaller than one molecular length at a temperature
still lower than $T_A$, the precise calculated value of the barrier at
this point should not be taken too seriously. The theory nevertheless
predicts that at temperatures above $T_{cr}$ the $\alpha$-relaxation
barriers have reached their lowest possible value, and a collisional
mechanism becomes the main contributor to the liquid's viscosity.  On
the other hand, the barrier renormalization becomes less significant
as one approaches $T_K$, the softening effects should vanish, subduing
themselves to the singularity at $T_K$. At $T_g$, however, this
renormalization is still quite noticeable, as represented by the
relative barrier value less than one. This renormalization coefficient
also has additional significance in that it tells one how the apparent
fragility of a given substance would differ if one were to use the
softening anzatz from Eq.(\ref{F_r_interp}) as opposed to the bare
RFOT's Eq.(\ref{F_r}) (this in addition to the factor $T_g/T_K$ that
may arise depending on which functional form of $s_c(T)$ is used, as
mentioned earlier).

At this point in the discussion, judging from the four analyzed
substances, a pattern emerges that stronger substances tend to have a
larger value of $T_A/T_K$ ratio. As mentioned, this was explained in
\cite{HW} by a varying degree of bonding in the various liquids and
indeed conforms to the lore that stronger substances tend to form a
larger number of directional bonds per molecular unit.

\subsection{\em Das Glasperlenspiel}

We have already mostly discussed the implications of the appearance
and significance of the temperature $T_A$ for interpreting the
features seen in plot of $\log \eta(T)$. Let us now check whether the
predictions of the entropy per bead needed for RFOT fits actually
matches our structural and thermodynamic expectations (the results of
the following discussion are summarized in Table \ref{bead_table}).
We phrase the issue of comparing microscopic theory and experiment in
this way because of the manner of the fit we have used and possible
ambiguities in defining $s_c$ (per bead). One could equally well argue
that it is the bare surface tension argument that is in error in the
RFOT theory so the surface tension should be modified. In fact
variations of less than 50\% in the bare surface tension give nearly
equivalent effects to changes in $s_c$. In order to conduct this
comparison of microscopic theory and experiment, we must determine the
number of beads for each material. First, there is a procedure based
on an intuitive parsing of the molecular structure: salol consists of
two benzene rings connected by a carbonyl and an oxygen - hence it
should have 4 beads (a benzene ring is a rigid entity). TNB
(tri-naphthyl benzene) consists of three naphthyl groups (effectively
double benzene rings) connected by a benzene - giving a count of 7
beads (we count a naphthyl as having two beads in spite of its
rigidity because it has a rotational degree of freedom around the
out-of-plane symmetry axis). Propanol, we would say, has three beads
(the OH group is hydrogen bonded to a neighbor).  Silica is a
difficult case because it is hard to estimate the degree of
independence of silicon and oxygen atoms in this highly networked
substance which some may consider a molten salt. Notwithstanding this
ambiguity, we assume here one bead per molecule (there is indeed one
oxygen per tetrahedron, which is, apropos, often thought to be the
moving unit in a-SiO$_2$ \cite{Trachenko}). As seen from Table
\ref{bead_table}, the results are quite reassuring, with the largest
disagreement between molecular structural intuition and RFOT theory
with softening being 57\% for salol.  Remember, there are at least
four independent sources of error potentially contributing to this
discrepancy, as discussed earlier.

\begin{table}
\small
\centerline{\begin{tabular}{|c||c||c|c||c|c||c|c|} 
\hline
substance  & $b_K \equiv s_{\infty}/s_{fit}$ & $b_{chem}$  & $b_K/b_{chem}$ & 
                         $b_{HS}$ & $b_K/b_{HS}$ & 
                         $b_{LJ}$ & $b_K/b_{LJ}$
\\ \hline \hline
Salol      & 6.29  & 4  & 1.57 & 6.32 & 1.00 & 4.36 & 1.44
\\ \hline
TNB        & 9.37  & 7  & 1.34 & 9.33 & 1.01 & 6.44 & 1.46
\\ \hline
1-Propanol & 2.95  & 3  & 0.98 & 4.20 & 0.70 & 2.90 & 1.02
\\ \hline
a-Silica   & 0.78  & 1  & 0.78 & 0.57 & 1.37 & 0.39 & 1.99
\\ \hline  
\end{tabular}}
\caption{\small The first column repeats the last column of Table
\ref{parmtr_table}, this would be the exact bead count for these
substances, if both $s_{\infty}$ and $s_{fit}$ were determined
precisely. Next six columns are organized into three pairs.  In each
pair, the first number gives the bead number estimated by a particular
method (see text) and the resultant ratio of the heat capacity jump at
$T_K$ per bead determined from this bead count to the same quantity as
obtained in the fit. $b_{HS}$ denotes the bead number per molecule
calculated using the entropy of melting of a hard sphere liquid
$s_{HS} = 1.16 k_B$.  $b_{LJ}$ is the same quantity but estimated
using the entropy of melting of a Lennard-Jones liquid $s_{LJ} = 1.68
k_B$.}
\label{bead_table}
\end{table}

\begin{table}
\small
\centerline{\begin{tabular}{|c||c|c|c|c|} 
\hline
substance  & $s_{c,T_g}^{theor}$,$k_B$  & $s_{c,T_g}^{exp}$(chem),$k_B$ & 
                              $s_{c,T_g}^{exp}$($HS$),$k_B$ & 
                              $s_{c,T_g}^{exp}$($LJ$),$k_B$
\\ \hline \hline
Salol      & 0.54 & 0.85 & 0.54 & 0.78
\\ \hline
TNB        & 0.56 & 0.75 & 0.56 & 0.81
\\ \hline
1-Propanol & 0.74 & 0.73 & 0.52 & 0.75
\\ \hline
a-Silica   & 0.78 & 0.61 & 1.0  & 1.54
\\ \hline  
\end{tabular}}
\caption{\small The first column lists the values of configurational
entropy per bead at the glass transition temperature $T_g$ as derived
from our fits: $s_c(T_g) = s_{fit} (1-T_K/T_g)$. These are to be
compared to the experimental values of $s_c^{exp}$ at $T_g$ per bead
obtaned according to $s_c^{exp}(per \:\: bead) = s_\infty
(1-T_K/T_g)/b$ using a bead count $b$ estimated by a particular method
(see text). A useful reference value of $s_c$ is $0.82$, predicted to
hold universally at $T_g$ on the one hour time scale by the RFOT
theory without softening.}
\label{scTg_table}
\end{table}

Now, admittedly, the bead assignment procedure has an element of
subjectivity. (Ways of presenting the thermodynamics near $T_g$ in a
bead count independent fashion have been used before (see for example
\cite{HuangMcKenna,WangAngell}). We may therefore propose a completely
automatized, interpretation free procedure, that uses measurable data
on the entropy of {\em crystallization}. The motivation for this is
that, clearly, a similar issue of bead counting would arise if we
tried to describe freezing of a molecular liquid by comparison with a
theory for atomic entities that are spherical.  Fortunately for us in
making this comparison, all four substances exist in crystalline form
(actually, more than one crystalline form for silica, where we will
use the low $T$ quartz polymorph). In the RFOT theory, the precise
meaning of a bead is that it should effectively behave as a hard
sphere in terms of its entropy loss in the ideal localized state at
$T_K$ and as the elementary unit of a droplet-droplet interface
\cite{XW}. From the entropy point of view a similar localization also
occurs at freezing. We may therefore estimate the number of (hard
sphere) beads in a particular molecule by dividing the entropy of
fusion of this substance by the entropy of fusion of a hard sphere
(HS) liquid, known to be $s_{HS} = 1.16 k_B$ \cite{Hansen}, that is
bead\#$=(\Delta H/T_m)/s_{HS}$.  Obviously, crystallization and
vitrification are different in several ways. For example, the freezing
occurs with a volume change ($\sim$10\% for a hard sphere liguid)
while the glass transition does not have a volume change. Thus
freezing will have a larger dependence on the attractive
forces. Hence, an alternate bead count may also take into account the
attractive force changes (the Lennard-Jones (LJ) liquid's entropy of
fusion is larger than that of a hard sphere liquid. We assume it is
equal to the fusion entropy of argon $s_{LJ} = 1.68 k_B$, which is
indeed a nearly LJ substance \cite{Rice_PCh}); similarly, the softness
of the core potential for metals enters into their fusion
entropies. We should not be surprised to find therefore a deviation
from the hard sphere estimate, but for chemically similar systems
there will be a more or less systematic deviation.  Notwithstanding
these caveats, the procedure now has the merit of being totally
unambiguous.  The results of counting beads by using both the freezing
entropy of a hard sphere and a Lennard-Jones liquid are also presented
in Table \ref{bead_table}. We see that our earlier conclusions still
hold, at least for these four substances. We note it is important that
both the glass and crystal have no remaining degrees of freedom
i.e. freezing does not produce plastic orienationally disordered
crystals.

Still, can we independently verify that using bead\# equal to $(\Delta
H/T_m)/s_{HS}$ is reasonable more generally? We have found several
reassuring signs that the answer is affirmative. Consider plotting the
kinetically estimated fragility index $m$ versus the following
quantity $m_{calc} = c T_g \Delta C_p/\Delta H_m$, as done by Wang and
Angell in \cite{WangAngell}. $\Delta C_p$ and $\Delta H_m$ are heat
capacity jump at $T_g$ and $\Delta H_m$ is the fusion enthalpy, both
per mole, not per bead. Clearly, Wang-Angell plots successfully avoid
the issue of how many independently moving units a molecule should be
considered to have. Empirically, they observed a correlation between
the kinematic $m$ and $m_{calc}$ (Fig.1 of \cite{WangAngell}). If one
draws a line through the densest linear cluster on the $m$
vs. $m_{calc}$ graph (which appears to edge the total cluster),
numerical value 56 for the slope $c$ fits well. However values of $c$
up 70 seem to be still consistent within the scatter of the data, if
all points are given equal weight.  On the other hand, we can
qualitatively estimate the value of $m_{calc}$ as predicted by the
RFOT. Assume for the sake of argument that the VTF in the vanilla form
from RFOT of $\tau = \tau_0 \exp(32./s_c)$ (as follows from
Eq.(\ref{F_sc}) holds (this is indeed a good approximation for many
substances). Additionally, assume the empirical form of the
configurational entropy, that we have used throughout the paper: $s_c
= s_{\infty} (1-T_K/T)$ \cite{RichertAngell}. We have seen the latter
expression is indeed a very good approximation. The corresponding
(temperature dependent) heat capacity jump is then simply $\Delta c_p =
s_{\infty} (T_K/T)$.  Simple algebra shows that, with these
assumptions, $m \equiv T [d \log_{10} \tau(T)/d(1/T)] = \Delta
c_p^b(T_g) (32./s_c^2(T_g)) \log_{10} e$. With our proposed way to
estimate the bead count from the fusion entropy, the heat capacity per
bead $\Delta c_p^b$ is then $\Delta c_p^b = \Delta c_p (s_{HS}
T_m/\Delta H_m)$. Finally, within bare RFOT,
\begin{equation}
m = \frac{T_g \Delta c_p(T_g)}{\Delta H_m} 
\left\{s_{HS} (\log_{10} e) \frac{32.}{s_c^2(T_g)} \frac{T_m}{T_g} \right\},
\end{equation}
where we remind the reader $s_c(T_g)$ is a per bead quantity
corresponding to the effective hard sphere model. The ratio $T_m/T_g$
is almost universally 3/2. Using the value $s_c(T_g) = .82$ at the
hour scale and $s_{HS} = 1.16$, the numerical factor in the curly
brackets (corresponding to $c$) is $\simeq 36.$.  If one were to use
the Lennard-Jones fusion entropy $s_{LJ} = 1.66$ rather than $s_{HS}$
in this expression, the numerical factor $c$ predicted for the
Wang-Angell plot is about $52$, rather close to the Wang-Angell
value. Additionally, note that the real $s_c(T_g)$ tends to be a bit
smaller when softening effects are included. Finally, we note that in
applications it may be beneficial to remember that ratio $T_g \Delta
C_p/\Delta H_m$ depends only on the ratio of $T_m$ to $T_K$ (see the
caption to Table \ref{parmtr_table}):
\begin{equation}
\frac{T_g \Delta C_p(T_g)}{\Delta H_m} = \frac{T_K}{T_m-T_K}.
\end{equation}

It appears a useful exercise to assess the quality of both our fitting
method and the bead number assignment by looking at the values of the
configurational entropy at $T_g$ per bead as extracted from the
theoretically computed $s_{fit}$ and from the experimental number
$s_{\infty}$ divided by our {\it \`{a} priori} bead count (see Table
\ref{scTg_table}). While all numbers appear to be reasonable (and in
fact comparable to the bare RFOT's $s_c(T_g) = 0.82$), only the
Lennard-Jones bead count for silica, again, seems too low, consistent
with our earlier findings.

In fact, we note, the comparison of glass physics with the fusion
entropy is quite venerable being intrinsic to Kauzmann's way of
plotting the configurational entropy as function of temperature,
divided by entropy loss $\Delta s_m$ during crystallization, again a
``bead'' independent quantity. At $T_g$, the plain RFOT and $s_{HS} =
1.16$ predicts $s_c/s_m \sim .82/1.16 \simeq .7$ (this falls within
the range of observed values. See a compendium of experimental values
of $s_c/s_m$ for polymers (both entropies per mole), given in
\cite{RSN}). As we have seen, with softening included, $s_c(T_g)$ is
not a constant and seems to range from $\sim .5$ to $\sim 1$ (see
Table \ref{scTg_table}).  Curiously the smallest value of this
quantity reported in \cite{RSN} is $\sim 0.1$ for PVC. We suspect this
exceptionally low apparent value of the configurational entropy at
$T_g$ may well be caused by partial crystallization that goes
unnoticed because it occurs locally and over a range of different
temperatures due to heterogeneity in local environment.

One observes that the values of $s_{fit}$ are apparently remarkably
close to each other. Therefore, the heat capacity jumps, as they would
be at $T_K$, will be much closer in magnitude than the apparent
experimental $\Delta c_p$'s at $T_g$, which differ at least by an
order of magnitude (note, our substances cover a good range of
fragilities that have been encountered so far). This notion supports,
albeit circumstancially, the usefullness of the concept of a ``bead''
as a reference motional unit in a glass. It seems, structural degrees
of freedom as expressed in terms of motions of the beads appear to
have much more in common, even quantitatively, in spite of apparent
chemical idiosyncrasies of different materials.  Nevertheless, nothing
in the theory so far insists that $\Delta c_p (T_K)$ per bead should
be the same for all substances. Remember, the size of the bead $a$
(that would affect the bead count), enters in the theory in two (what
so far seems) largely independent places. On the one hand, it is the
reference length scale in the argument on the surface tension
coefficient $\sigma_0$; on the other hand, it determines the volume
density of beads as derived from the configurational entropy data.

\begin{figure}[tbh]
\centerline{\psfig{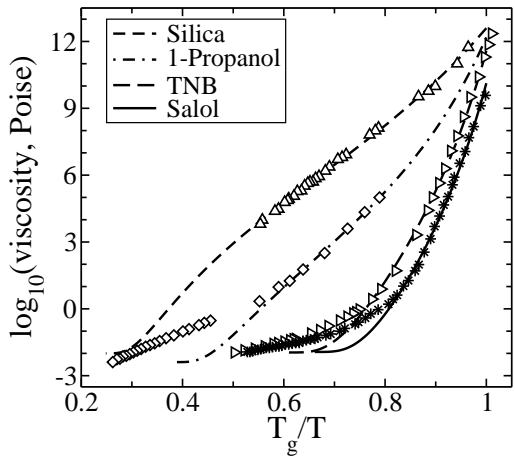}}
\caption{Shown is the Angell plot of the log-viscosity data as a
function of $T_g/T$. The symbols denote experimental points.  The
lines are our the RFOT results (shown only between $T_A$ and $T_g$).}
\label{Tg_T}
\end{figure}

\begin{figure}[tbh]
\centerline{\psfig{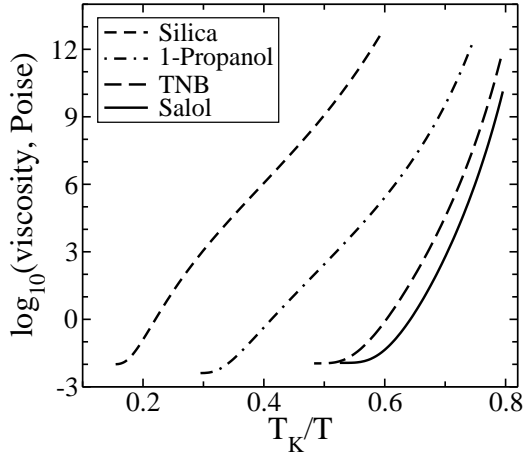}}
\caption{Theoretical log-viscosity curves as functions of $T_K/T$
(shown only between $T_A$ and $T_g$).}
\label{TK_T}
\end{figure}

One can appreciate some of these points graphically by plotting
$\log[\eta(T)]$ versus $T_g/T$ and $T_K/T$, as shown in
Figs.\ref{Tg_T} and \ref{TK_T} respectively.

Note that the easily visible difference in slopes between the curves
when viscosity is plotted vs. $T_g/T$, become rather less pronounced
in the $T_K/T$ graph.  According to the present theory, some of the
differences in the qualitative behavior of the $\log(\eta)$
temperature dependence are due to varying degree to which the
softening effects play a role at the glass transition. This degree is
directly related to the ratio of the two basic energy scales in the
theory: $T_A$ and $T_K$. According to RFOT, $T_K$ gives the energy
scale of the ideal glass transition, below which structural
equilibration is impossible in principle. $T_A$, on the other hand
gives a reference energy at which the liquid would fall out of
equilibrium, if activated events were eliminated as in mean-field
theory. The glass transition temperature $T_g$, in contrast, bears no
fundamental meaning, but only reflects the amount of patience of the
experimenter.  We suggest that plotting the log-viscosity data as a
function of $T_K/T$ (with thermodynamically derived $T_K$) will make
it easier to separate to some extent the phenomena characteristic of
the ideal glass transition from those occuring near the onset of
activated transport.

\section{Discussion and Conclusions}

The RFOT theory of supercooled liquid dynamics differs from a pure
"entropy" theory by exhibiting the phenomenon of barrier softening as
the spinodal or mode coupling temperature $T_A$ is approached from
below.  We have shown here that an approximate theory of these
softening effects indeed accounts for the deviations from the VTF law
that experimentalists have previously noticed.

From a formal point of view, the present treatment of the barrier
softening is not complete, especially by not being symmetric about
$T_A$.  Above $T_A$, if the naive MCT relaxation time is sufficiently
long, activated events can short circuit the predicted sigularly slow
relaxation from MCT. Likewise, additional mode coupling effects may
play a role just below $T_A$. Treating this feature is a problem for
future work.  We feel that it concerns, on a logarithmic scale, only a
small part of the dynamic range from a few centipoise to a few
decapoise, a small fraction of the range of viscosities explored in
the laboratory.  This is however the range most easily addressed by
computer simulation currently so further work on the problem is
warranted.

We also comment that in this paper we treat two quantities that enter
the theory as independent, one $\Delta c_p$ (at $T_g$) and the other
the ratio $T_A/T_K$.  Yet the curve fits show these two quantities
clearly are correlated (empirically so far).  Indeed in some detailed
microscopic models the quantities are correlated too.  Hall and
Wolynes' calculations suggest they are both functions of the fraction
of bonds that are made in a model network liquid.  Yet the
Hall-Wolynes treatment suggests they should also be functions of the
detailed force laws as well. Thus there may be ways of teasing the two
features determining fragility apart by a judicious choice of systems
e.g. liquid metals.  It is also possible that the correlation may
arise from some common mechanism of frustration in many glasses, like
icosahedral order formation in flat three dimensional space. In this
case finding systems that break the correlation in the laboratory will
prove difficult.

As to the molecular structural details it is perhaps remarkable that
the RFOT theory does as well as it does.  Indeed we see RFOT theory
encounters strong parallels with the difficulties faced in the
molecular theory of crystallization.  Perhaps some of the complexities
of that problem, polymorphisms and residual ordering transitions in
the solid state (plastic crystallinity) are ameliorated here because
they actually enter in only a statistical rather than a state specific
way for glasses.  In any event we note that the assumption of hard
sphere geometry is by no means required by the RFOT theory approach
which can make use of density functional methods to treat molecular
network systems as has already been done to some extent.  Such more
detailed free energy functional models can be used to predict barriers
by again constructing spatially inhomogeneous solutions of the
appropriate molecular density functional corresponding with entropic
droplets.

The barrier softening effects on the predicted length scales of
dynamical heterogeneity are very modest near $T_g$. Because of this
those predictions of the RFOT theory that are sensitive to length
scale like the degree of non-exponentiality \cite{XWbeta} and density
of the two-level systems are only moderately affected.  Indeed since
the lengths are a bit smaller that the vanilla RFOT, the Kohlrausch
$\beta$ correlation between theory and experiment should improve (i.e.
less dispersion of relaxation times). The lengths do decrease faster
as $T$ increases from $T_g$ than they do in vanilla RFOT theory and
this is consistent with the temperature dependence of violations of
the Stokes-Einstein relation recently measured in TNB
\cite{Ediger_hydro}. Indeed the RFOT calculation with softening
predicts no correction above the break temperature $T_{cr}$, in
agreement with experiment.

In summary, this work has furthered the quantitative development of
the random first order transition theory of the glass transition:
First, we have computed how the activation barriers for structural
relaxation in supercooled liquids are modified in the regime when the
liquid is only marginally supercooled, that is near to the so called
dynamical transition temperature $T_A$. $T_A$ is in fact the
temperature where alternative metastable liquid packings begin having
statistical significance, as the liquid is cooled.  Fluctuations of
the order parameter at this dynamical transition at $T_A$ lead to
vanishing surface tension between distinct aperiodic structures into
which the liquid packs itself locally when supercooled. We call this
droplet interface softening. As a result of this calculation, we have
understood the significance of two distinct regimes in the relaxation
time temperature dependence.  The ``diverging'' low temperature part,
represented by liquid's viscosity above $\sim$10 poise, indeed
corresponds to local cooperative rearrangements, microscopically
understood by the random first order transition calculation. Molecular
transport in the higher $T$ portion, on the other hand, is dominated
by a multiple collision mechanism, described by the mode-mode coupling
theory, though strongly affected by the presence of activation, which
is very facile at this point. Usually, the high temperature segment of
the $\log \tau(T)$ plot has been difficult to fit with the empirical
VTF functional form. The theory makes clear that different mechanisms
are responsible for molecular transport in systems at these different
temperatures, as just explained. Our fitting of experimental data
gives an unambiguous answer for the value of temperature $T_{cr}$ at
which the crossover between the two regimes occurs. It also yields the
value of $T_A$.  We use the value of excess configurational entropy
derived from calorimetry as input in the calculation. Finally, we
compute the cooperativity length scales for structural relaxation in
supecooled melts.  While these length scales are around 5-6 molecular
spacings at the glass transition, as predicted by the RFOT even
without softening effects, they rapidly decrease upon warming and
vanish at temperatures above $T_{cr}$.

{\em Acknowledgments:} We are grateful to Xiaoyu Xia for lively
discussions on this problem. The work of PGW in this area is supported
by NSF grant CHE 0317017.


\begin{thebibliography}{99}

\bibitem{KW} T.R.Kirkpatrick and P.G.Wolynes,
Phys. Rev. B, {\bf 36} 8552 (1987).


\bibitem{KTW} T.R.Kirkpatrick, D.Thirumalai and P.G.Wolynes, 
Phys. Rev. A, {\bf 40} 1045 (1989).
 
\bibitem{XW} X.Xia and P.G.Wolynes, PNAS, {\bf 97}, 2990 (2000);
cond-mat/9912442.

\bibitem{Spiess} U. Tracht, M. Wilhelm, A. Heuer, H. Feng,
K. Schmidt-Rohr, and H. W. Spiess Phys. Rev. Lett. {\bf 81}, 2727
(1998).

\bibitem{Israeloff} E.V.Russell and N.E.Israeloff,
Nature, {\bf 408} 695 (2000).

\bibitem{AdamGibbs} G.Adam and J.H.Gibbs, J. Chem. Phys.
{\bf 43}, 139 (1965).

\bibitem{EastwoodW} M.P.Eastwood and P.G.Wolynes, Europhys. Lett., 
{\bf 60} 587 (2002).

\bibitem{Villain} J.Villain, J. Physique {\bf 46}, 1843 (1985).

\bibitem{XWbeta} X.Xia and P.G.Wolynes, Phys. Rev. Lett.
{\bf 86}, 5526 (2001); cond-mat/0008432.

\bibitem{XWhydro} X.Xia and P.G.Wolynes,
J. Phys. Chem. B {\bf 105} 6570 (2001); cond-mat/0101053.

\bibitem{CiceroneEdiger} M.T.Cicerone and M.D.Ediger,
J. Chem. Phys. {\bf 104}, 7210 (1996). 

\bibitem{Ediger_hydro} S.F.Swallen, P.A.Bonvallet, R.J.McMahon, and M.D.Ediger,
Phys. Rev. Lett. {\bf 90}, (2003) 015901.

\bibitem{LW} V.Lubchenko and P.G.Wolynes, Phys. Rev. Lett.
{\bf 87}, 195901 (2001).

\bibitem{LW_BP} V.Lubchenko and P.G.Wolynes, Proc. Natl. Acad. Sci. USA
{\bf 100}, 1515 (2003).

\bibitem{Lindemann} F.A.Lindemann, Phys. Z., {\bf 11}, 609 (1910).

\bibitem{Kauzmann} W.Kauzmann, Chem. Rev. {\bf 43}, 219 (1948).

\bibitem{SGandB} M.Mezard, G.Parisi and M.A.Virasoro, ``Spin Glasses
and Beyond'', World Scientific, 1987.

\bibitem{Nelson} D.R.Nelson, ``Defects and Geometry in Condensed
Matter Physics'', Cambridge University Press, 2002.

\bibitem{Nussinov} Z.Nussinov, cond-mat/0209292.

\bibitem{SchmalianW} J.Schmalian, P.G.Wolynes, Phys. Rev. Lett., {\bf
85}, 836 (2000).

\bibitem{Mezei} F.Mezei, in ``Liquids, Freezing and the Glass
Transition'', ed. J.P.Hansma, D.Levesque and J.Zinn-Justin,
Nort-Holland, Amsterdam, p.629, 1991.

\bibitem{DasguptaValls} C.Dasgupta and O.T.Valls, Phys. Rev. E {\bf
59}, 3123 (1999).

\bibitem{PGWaper} P.G.Wolynes, in Proceedings International
Symposium on Frontiers in Science (Hans Frauenfelder Festschrift),
S. Chan and P. G. DeBrunner, Editors (Am. Inst. Physics, 1989).

\bibitem{HeniLowen} M.Heni and H.L\"{o}wen, Phys. Rev. E, {\bf 60}
7057 (1999).

\bibitem{Laird} B.B.Laird, J. Chem. Phys., {\bf 115}, 2887 (2001).

\bibitem{Turnbull} D.Turnbull, J. Appl. Phys. {\bf 21}, 1022 (1950).

\bibitem{Leutheusser} E.Leutheusser, Physical Review A, {\bf 29} 2765 (1984).

\bibitem{Gotze_MCT} U.Bengtzelius, W.G\"{o}tze, A.Sjolander,
J. Phys. C: Solid State Physics, {\bf 17} 5915 (1984).

\bibitem{Kirkpatrick} T.R.Kirkpatrick, Phys. Rev. A, {\bf 31} 939
(1985).

\bibitem{SSW} Y.Singh, J.P.Stoessel, and P.G.Wolynes,
Phys. Rev. Lett. {\bf 54}, 1059 (1985). 

\bibitem{SW} J.P.Stoessel and P.G.Wolynes,
J. Chem. Phys. {\bf 80}, 4502 (1984).



\bibitem{RamakrishnanYussouff} T.V.Ramakrishnan and M.Yussouff,
Phys. Rev. B {\bf 19}, 2775 (1979); M.Yussouff, Phys. Rev. B {\bf 23}, 
5871 (1983).

\bibitem{MezardParisi} M.Mezard, G.Parisi, J. Chem. Phys., {\bf 111} 1076 (1999);

\bibitem{KW_A} T.R.Kirkpatrick and P.G Wolynes, Phys. Rev. A, {\bf 35}
3072 (1987).

\bibitem{Widom} J.S.Rowlinson and B.Widom, ``Molecular Theory
of Capillarity'', Clarendon Press, Oxford, 1982.

\bibitem{Feynman} R.P.Feynman, Phys. Rev. {\bf 94}, 262 (1954).

\bibitem{deGennes} P.G. de Gennes, ``Scaling Concepts
in Polymer Physics'', Cornell University Press, 1979.

\bibitem{Binder} C.Billotet and K.Binder, Z. Physik B,
{\bf 32} (1978) 195; and references therein.

\bibitem{Glotzer} C.Donati, S.Franz, S.C.Glotzer, and G.Parisi,
J. Non-Cryst Solids, {\bf 307}, 215 (2002).

\bibitem{Parisi} G.Parisi, Nuovo Cimento {\bf 16D}, 939 (1994).

\bibitem{TakadaW} S. Takada and P. G. Wolynes, Phys Rev. E. 55, 4562 (1997)

%\bibitem{FisherBerker} M.E.Fisher and A.N.Berker,
%Phys. Rev. B, {\bf 26} (1982) 2507.

\bibitem{HuangMcKenna} D.Huang and B.McKenna, J. Chem. Phys.
{\bf 114}, 5621 (2001).

\bibitem{MartinezAngell} L.-M.Martinez and C.A.Angell,
Nature {\bf 410}, 663 (2001).

\bibitem{Mohanty} U.Mohanty, N.Craig and J.T.Fourkas, J. Chem. Phys.
{\bf 114}, 10577 (2001).

\bibitem{RichertAngell} R.Richert and C.A.Angell, J. Chem. Phys.
{\bf 108}, 9016 (1998).

\bibitem{Magill} J.H.Magill, J. Chem. Phys.
{\bf 47}, 2802 (1967).

\bibitem{Richet} A.Sipp, Y.Bottinga and P.Richet,
  J. Non-Cryst. Solids, {\bf 288}, 7 (2001).

\bibitem{Stickel2} F.Stickel, E.W.Fischer and R.Richert,
J. Chem. Phys. {\bf 104}, 2043 (1996).

\bibitem{TNBexp} D.J.Plazek and J.H.Magill, J. Chem. Phys.  {\bf 49},
3678 (1968).

\bibitem{RSN} C.M.Roland, P.G.Santangelo, and K.L.Ngai,
J. Chem. Phys. {\bf 111}, 5593 (1999).

\bibitem{Stickel} F.Stickel, E.W.Fischer and R.Richert,
J. Chem. Phys. {\bf 102}, 6251 (1995). This paper also
describes fits of salol's relaxation data using several 
alternative theories.

\bibitem{Kivelson} D.Kivelson, S.A.Kivelson, X.L.Zhao, Z.Nussinov, and
G.Tarjus, Physica A, {\bf 219}, 27 (1995).

\bibitem{HorbachKob} J.Horbach and W.Kob, Phys. Rev. B {\bf 60}, 3169
(1999).

\bibitem{mAngell} R.B\"{o}hmer and C.A.Angell, Phys. Rev. B {\bf 45},
10091 (1992).

\bibitem{mNgai} D.J.Plazek and K.L.Ngai, Macromolecules {\bf 24},
1222 (1991).

\bibitem{Schweizer} K.S.Schweizer and E.J.Saltzman,
J. Chem. Phys. {\bf 119}, 1181 (2003).

\bibitem{Roland} R.Casalini, M.Paluch, and C.M.Roland,
J. Chem. Phys. {\bf 118}, 5701 (2003).

\bibitem{Ngai_rev} K.L.Ngai, J. Non-Cryst. Solids, {\bf 275}, 7
(2000).

\bibitem{HW} R.W.Hall and P.G.Wolynes, Phys. Rev. Lett. {\bf 90},
085505 (2003).

\bibitem{Trachenko} K.Trachenko, M.T.Dove, M.J.Harris, and V.Heine,
J. Phys.: Condens. Matter, {\bf 12} 8041 (2000).

\bibitem{Rice_PCh} R.S.Berry, S.A.Rice and J.Ross, ``Physical
Chemisry'', Secind Edition, Oxford University Press, 2000.

\bibitem{WangAngell} L.-M.Wang and C.A.Angell, J. Chem. Phys. {\bf 118}, 10353 
(2003).

\bibitem{Hansen} J.P.Hansen and I.R.McDonald, ``Theory of Simple Liquids'',
Academic Press, New York, 1976.


\end{thebibliography}
\end{document}